\documentclass[aps,prb,floatfix,twocolumn,showpacs]{revtex4}
\usepackage{epsfig,amsmath,amssymb}

\begin{document}

%\twocolumn[
\hsize\textwidth\columnwidth\hsize\csname@twocolumnfalse\endcsname

\title{Spin properties of single electron states in coupled quantum dots}

\author{Peter Stano$^{1,2}$ and Jaroslav Fabian$^{1,2}$}
\affiliation{$^1$Institute of Physics, Karl-Franzens University,
Universit\"atsplatz 5, 8010 Graz, Austria \\ $^2$Institute for Theoretical 
Physics, University of Regensburg, 93040 Regensburg, Germany}

\vskip1.5truecm
\begin{abstract}
Spin properties of single electron states in laterally coupled quantum dots in the 
presence of a perpendicular magnetic field are studied by exact
numerical diagonalization. Dresselhaus (linear and cubic) and Bychkov-Rashba spin-orbit couplings are 
included in a realistic model of confined dots based on GaAs. Group theoretical 
classification of quantum states with and without spin orbit coupling is
provided. Spin-orbit effects on the g-factor are rather weak.
It is shown that the frequency of coherent oscillations (tunneling
amplitude) in coupled dots is largely unaffected by spin-orbit effects 
due to symmetry requirements. The leading contributions to the frequency
involves the cubic term of the Dresselhaus coupling. Spin-orbit coupling
in the presence of magnetic field leads to a spin-dependent tunneling
amplitude, and thus to the possibility of spin to charge conversion, namely
spatial separation of spin by coherent oscillations in a uniform magnetic field. 
It is also shown that spin hot spots exist in coupled GaAs dots already at 
moderate magnetic fields, and that spin hot spots at
zero magnetic field are due to the cubic Dresselhaus term only.
\end{abstract}
\maketitle

\section{Introduction}

The possibility of tuning spin-orbit coupling 
\cite{datta1990:APL,nitta1997:PRL, koga2002:PRL} in low-dimensional 
semiconductor electronic structures has stirred great interest in spin properties 
of lateral semiconductor electron systems in the presence of Dresselhaus \cite{dresselhaus1955:PR} 
and Bychkov-Rashba \cite{bychkov1984:JPC,rashba1960:FTT} spin-orbit couplings. 
The former appears in low-dimensional systems lacking inversion symmetry
in the bulk (such as zinc-blende semiconductors), 
the latter in low-dimensional structures with asymmetric confining potentials.
The principal question is what spin and charge properties and to what degree can be
affected and manipulated by this tuning. Such questions are of fundamental
importance for spintronics.\cite{zutic2004:RMP}  

Electron spins in coupled quantum dot systems have been proposed to perform universal
gating of quantum computers. \cite{loss1998:PRA}
The spin acts as a qubit and exchange coupling provides the physical 
realization of two-qubit gates. \cite{burkard1999:PRB,hu2000:PRA}
Another application of a controlled coupling between dots is spin entanglement distillation
in which singlet and triplet states get spatially separated during
adiabatic passage through trapped states. \cite{fabian2004:CM}
Understanding of spin-orbit properties of coupled dots is thus of
great interest to quantum information processing. 

Spin-orbit coupling provides a way for orbital degrees of freedom
to influence spin states. As a result the spin dynamics is affected, making
spin qubit operations more complex
(it was shown, though, that two qubit operations can be performed
reliably even in the presence of spin-orbit interaction which 
leads to anisotropic exchange\cite{burkard2002:PRL,stepanenko2003:PRB}).
Furthermore, spin-orbit coupling leads to spin decoherence and relaxation 
due to phonons, \cite{khaetskii2001:PRB, destefani2004:CM, woods2002:PRB, cheng2004:PRB,golovach2004:PRL,bulaev2005:PRB}
limiting the operation time. The impressive experimental progress in coherent oscillations in coupled dot 
systems \cite{petta2004:PRL, hayashi2003:PRL, wiel2003:RMP, petta2004:CM}, 
as well in spin dephasing and spin manipulation in single
\cite{hanson2003:PRL, hanson2005:PRL} and double dots \cite{johnson2005:CM},
provides additional strong impetus for investigating spin states
in double dots. Spin-orbit properties of single dots have been already
extensively investigated.
\cite{voskoboynikov2001:PRB,governale2002:PRL,sousa2003:PRB,rodriguez2004:PRB, destefani2004b:PRB,destefani2004:PRB,kuan2004:JAP,tsitsishvili2004:PRB,voskoboynikov2003:JAP,bulgakov2001:JETP,pfeffer1999:PRB,destefani2005:PRB}

In this paper we investigate the role of spin-orbit coupling, represented by
the Dresselhaus (both linear and cubic) and Bychkov-Rashba terms, in spin 
and charge properties of two laterally coupled quantum dots based on GaAs 
materials parameters. We perform numerically exact calculations of the energy
spectrum using the method of finite differences. We first study the general  
structure of the energy spectrum and the spin character of the states 
of the double dot system. We
construct the group theoretical correlation diagram for the single and
double dot states and indicate the possible transitions due to spin-orbit
coupling. This group theoretical classification is used in combination with
L\"owdin perturbation theory to explain analytically our numerical results.
In particular, we show that while allowed by symmetry, the specific forms
of the linear spin-orbit interactions do not lead to spin hot spots in the absence
of magnetic field (spin hot spots are nominally degenerate states lifted
by spin-orbit coupling\cite{fabian1998:PRL}). Only the cubic Dresselhaus term gives spin hot spots. If identified experimentally, the strength of the cubic term can be detected.

We next focus on two important measurable parameters: electronic
g-factor and tunneling amplitude. In single dots the variation of the effective
g-factor with the strength of the spin-orbit interaction has been investigated
earlier.\cite{sousa2003:PRB} The effect is not large, amounting to a 
fraction of a percent. Similar behavior is found for double dots. In 
our case of GaAs the contribution to the
g-factor from spin-orbit coupling is typically about 1\%, due to 
the linear Dresselhaus term.

More exciting is the prospect of influencing coherent tunneling oscillations
between the dots by modulating the spin-orbit coupling strength. Two effects can
appear: (i) the tunneling amplitude or frequency can be modulated by 
spin-orbit coupling and, (ii) the tunneling amplitude can be spin dependent.
We show how a naive application of perturbation theory leads to a misleading
result that (i) is present in the second order in linear spin-orbit coupling 
strengths, giving rise to an effective tunneling Hamiltonian involving  
spin-flip tunneling at zero magnetic field. Both numerical calculations and an analytical argument,
presented here, show that this is incorrect and that there is no correction
to the tunneling Hamiltonian in the second order of linear spin-orbit terms.
The dominant correction in the second order comes from the interference of 
linear and cubic Dresselhaus terms. We propose to use this criterion,
that the corrections to linear terms vanish in the second order, to distinguish
between single and double dots as far as spin properties of the states are
concerned. Indeed, at very small and very large intradot couplings the states
have a single dot character and the correction to energy due to linear
spin orbit terms depends on the interdot distance (except for the two lowest
states which provide tunneling).  We find that dots are ``coupled'' up to 
the interdot distance of about five single-dot confinement lengths.

In the presence of magnetic field the time reversal symmetry is broken. The
presence of spin-orbit coupling then in general leads to a spin dependent 
tunneling amplitude.  Spin up and spin down states will
oscillate between the two dots with different frequencies 
(for our GaAs dots the relative difference of the frequencies is at the
order of 0.1\%, but is higher in materials with larger spin-orbit
coupling). This leads to a curious physical effect, namely, that of
a spatial separation of different spin species. Indeed, starting with an 
electron localized on one dot, with a spin polarized in the plane (that is,
a superposition of up and down spins), after a sequence of coherent oscillations
the electron state is a superposition of spin up localized on one, and spin
down localized on the other dot. A single charge measurement on one dot collapses
the wave function to the corresponding spin state, realizing a spin to charge
conversion. We construct an effective, four state (two spin and two sites) 
tunneling Hamiltonian for the single electron double dot system, which takes 
into effect the above results. Such a Hamiltonian should be useful for 
constructing realistic model theories of spin dephasing, spin tunneling, and
kinetic exchange coupling in coupled quantum dot systems.

The paper is organized as follows. In Sec. II we introduce the model, the 
numerical technique, and materials and system parameters. In Sec. III we review 
the benchmark case of single dots with spin-orbit coupling and magnetic
field. Coupled double dots are studied in Sec. IV, separately in zero 
and finite magnetic fields. We conclude with the discussion of our results
in Sec. V.

\section{Model}
We consider conduction electrons confined in a [001] plane of a zinc-blende
semiconductor heterostructure, with additional confinement into lateral
dots given by appropriately shaped top gates. A magnetic
field $\textbf{B}$ is applied perpendicular to the plane. In the effective
mass approximation the single-electron Hamiltonian of such a system, 
taking into account spin-orbit coupling, can be decomposed into several terms:
\begin{equation} \label{eq:hamiltonian}
H=T+V_C+H_Z+H_{BR}+H_{D}+H_{D3}. 
\end{equation}
Here $T=\hbar^2\textbf{K}^2/2m$ is the kinetic energy with the effective
electron mass $m$ and kinetic momentum $\hbar\textbf{K} = \hbar\textbf{k} + e\textbf {A} = 
-i\hbar{\nabla}+e\textbf{A}$;  $e$ is the proton
charge and  $\textbf{A}=B(-y/2,x/2,0)$ is the vector potential of $\textbf{B} = (0,0,B)$. Operators of angular momentum with mechanical and canonical momenta are denoted as  $\textbf{L}=\textbf{r}\times(\hbar\textbf{K})$ and $\textbf{l}=\textbf{r}\times(\hbar\textbf{k})$.
The quantum dot geometry is described by the confining potential
$V_C(\textbf{r})$. Single dots will be described here by a parabolic potential
$V_C = (1/2) m \omega_0^2 r^2$, characterized by confinement energy 
$E_0 = \hbar\omega_0/2$ and confinement length $l_0 = (\hbar/m\omega_0)^{1/2}$,
setting the energy and length scales, respectively.  
Coupled double dots will be described by two displaced (along $\mathbf{x}$) parabolas:
\begin{equation} \label{eq:doubledot} 
V_C^{dd} = \frac{1}{2} m \omega_0^2 [(|x|- l_0 d )^2+y^2];
\end{equation}
the distance between the 
minima is $2d$ measured in the units of $l_0$.  
The Zeeman energy is given by $H_Z=(g^*/2)\mu_B\sigma_z B$, where
$g^*$ is the conduction band g-factor, $\mu_B$ is the Bohr magneton, 
and $\sigma_z$ is the Pauli matrix. In order to relate the magnetic moment
of electrons to their orbital momentum, we will use dimensionless parameter
$\alpha_Z = g^* m/2 m_e$, where $m_e$ is the free electron mass. 

Spin-orbit coupling gives additional terms in confined systems.\cite{zutic2004:RMP}
The Bychkov-Rashba Hamiltonian,\cite{rashba1960:FTT, bychkov1984:JPC}
\begin{equation}
H_{BR}=\tilde{\alpha}_{BR}\left(\sigma_x K_y-\sigma_y K_x \right),
\end{equation}
appears if the confinement is not symmetric in the growth direction (here $\textbf{z}$).
The strength $\tilde{\alpha}_{BR}$ of the interaction can be tuned by modulating
the asymmetry by top gates. Due to the lack of spatial inversion symmetry in 
zinc-blende semiconductors, the spin-orbit interaction of conduction electrons 
takes the form of the Dresselhaus Hamiltonian \cite{dresselhaus1955:PR} which, 
when quantized in the growth direction $\textbf{z}$ of our heterostructure gives 
two terms, one linear and one cubic in momentum:\cite{dyakonov1986:FTP}
\begin{eqnarray}
H_D&=&\gamma_c \langle K_z^2\rangle\left(-\sigma_x K_x+\sigma_y K_y\right),\\
H_{D3}&=&(\gamma_c/2)\left(\sigma_x K_x K_y^2-\sigma_y K_y K_x^2\right)+H.c.,
\end{eqnarray}
where $\gamma_c$ is a material parameter. The angular brackets in $H_D$ denote quantum 
averaging in the $\textbf{z}$ direction--the magnitude of $H_D$ depends on the
confinement strength. We will denote the sum of the two linear spin-orbit terms by 
$H_\mathrm{lin} = H_D+H_{BR}$. The complete spin-orbit coupling is then 
$H_{SO}=H_\mathrm{lin}+H_{D3}$. 
We find it useful to introduce dimensionless strengths of the individual terms
of the spin-orbit interaction by relating them to the confinement energy of a single
dot $E_0$.
We denote $\alpha_{BR}=\tilde{\alpha}_{BR}/E_0 l_0$ and 
$\alpha_D=\gamma_c\langle k_z^2\rangle/E_0 l_0$ for linear terms,
and $\alpha_{D3}=\gamma_c/2 E_0l_0^3$ for the cubic Dresselhaus term.

In our numerical examples we use $E_0=\hbar \omega_0/2=1.43$ meV for the confinement energy, which corresponds to 
the confinement length of $l_0=20$ nm. 
We further use bulk GaAs materials parameters: 
$m=0.067\,m_e$, $g^*=-0.44$, and $\gamma_c=27.5$ eV$\mathrm{\AA^3}$. For $\langle k_z^2\rangle$ 
we choose $5.3\times 10^{-4}\mathrm{\AA^2}$, which
corresponds to $\gamma_c\langle k_z^2\rangle=14.6$ meV \AA. This value of $\langle k_z^2\rangle$ corresponds to the ground state of a 6 nm thick triangular potential well.\cite{sousa2003:PRB}
For $\tilde{\alpha}_{BR}$ we choose a generic value of 4.4 meV\AA, which is in line of experimental observations.\cite{miller2003:PRL, knap1996:PRB} The dimensionless parameter of the Zeeman splitting then is $\alpha_Z = -0.015$,
while the relative strengths of the spin-orbit interactions are 
$\alpha_{BR} \approx 0.015$, $\alpha_D\approx 0.05$, and $\alpha_{D3} \approx 0.001$. 
Except for anti-crossings, the spin-orbit interaction is a small perturbation to the electronic
structure; it is, however, essential for investigating spin properties.

Our analytical calculations will often refer to the Fock-Darwin 
\cite{fock1928:ZP, darwin1931:pcps} spectrum, which is the spectrum of Hamiltonian 
\eqref{eq:hamiltonian} for a single dot with $H_{SO}=0$.
The corresponding wave functions $\Psi$ (expressed in polar coordinates $r$ and $\phi$), and energies $\epsilon$ are 
\begin{eqnarray} \label{eq:fock darwin functions}
\Psi_{n,l,\sigma}(r,\phi)&\!\!\!\!=&\!\!\!\!C\rho^{|l|}e^{-\rho^2/2}L_n^{|l|}(\rho^2)e^{i
l\phi}\xi(\sigma),
\\\label{eq:fock darwin energies}
\epsilon_{n,l,\sigma}&\!\!\!\!=&\!\!\!\!2E_0\frac{l_0^2}{l_B^2}(2n+|l|+1)+B\frac{\hbar e}{2m}(l+\alpha_Z\sigma),
\end{eqnarray} 
where $\rho=r/l_B$ is the radius in the units of the
effective confinement length $l_B$, defined by $l_B^{2}=l_0^{2}/\surd(1+B^2e^2 l_0^4/4\hbar^2)$;
$n$ and $l$ are the principal and orbital quantum numbers, respectively, 
$C$ is the state dependent normalization constant, and $L_n^{|l|}$ are associated
Laguerre polynomials. Spinors $\xi(\sigma)$ describe the spin $\sigma$ state of the
electrons. Since the parabolic dot has rotational
symmetry in the plane, the states have well defined orbital momentum $l$ and 
spin $\sigma$ in the z-direction. We also introduce a useful dimensionless measure $\theta$ of
the strength of the magnetic field induced confinement compared to 
the potential confinement: $\theta=B e l_B^2/2\hbar$, $0 < \theta< 1$. The parameter $\theta$ gives the number of magnetic flux quanta through a circle with radius $l_B$. For large magnetic fields $\theta\approx 1-(Bel_0^2/2\hbar)^2/2$. The confining length can be expressed as $l_B=l_0 (1-\theta^2)^{1/4}$.  

As it is not possible to solve for the spectrum of Hamiltonian \eqref{eq:hamiltonian}
analytically, we treat it numerically with the finite differences method using 
Dirichlet boundary conditions (vanishing of the wave function at boundaries).
The magnetic field is included via the Peierls phase: if $H(\textbf{r}_i,\textbf{r}_j)$ 
is the discretized Hamiltonian connecting grid points $\textbf{r}_i$ and $\textbf{r}_j$
at $B=0$, the effects of the field are obtained by adding a gauge phase: 
$H(\textbf{r}_i,\textbf{r}_j)\exp[i(e/\hbar)\int_{\textbf{r}_i}^{\textbf{r}_j}\textbf{A}.\mathrm{d}\textbf{l}]$. In our simulations we typically use $50\times50$ grid points. 
The resulting matrix eigenvalue problem is solved by Lanczos diagonalization.
The achieved accuracy is about $10^{-5}$. 

\section{Single dots}
\begin{figure} 
\centerline{\psfig{file=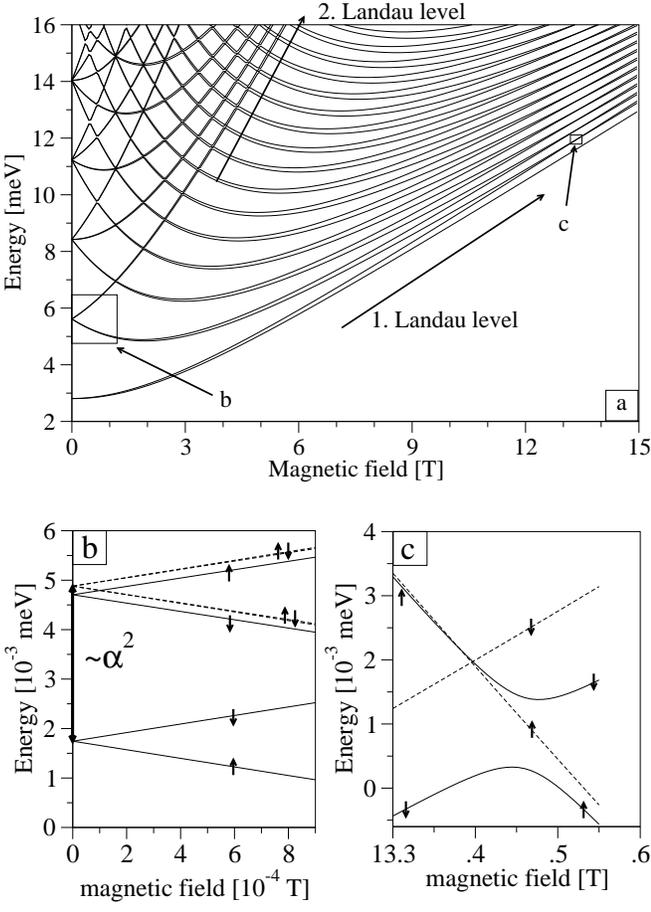,width=1\linewidth}}
\caption{Energy spectrum of a single dot in magnetic field.  a) The Fock-Darwin spectrum, Eq. \eqref{eq:fock darwin energies}. b)
Lowest orbital excited levels ($n=0$, $|l|=1$) without (dashed) and with (solid) spin-orbit coupling. 
Arrows indicate the spin states. For clarity the energy's origin here is shifted relative to case a). 
Both the shift in energy levels as well as the splitting at $B=0$ grow as $\alpha^2$. c) Anti-crossing
at the critical magnetic field (here about 13 T). For clarity, a linear trend was subtracted from the
data. }
\label{fig:fock darwin spectrum}
\end{figure}

As a starting point we review the effects of spin-orbit coupling in single dots. We are interested
in the changes to the spectrum and, in particular, to the magnetic moment of the lowest states, 
that is, to the effective $g$-factor. The calculated spectrum of a single dot is shown in 
Fig. \ref{fig:fock darwin spectrum}. There are three ways in which spin-orbit coupling influences the spectrum: (i) First,
the levels are uniformly shifted down, in proportion to $\alpha^2$ (by $\alpha$ here we mean any of
$\alpha_{BR}$, $\alpha_D$, or $\alpha_{D3}$). (ii) Second, the degeneracy at $B=0$ is lifted, again in
proportion to $\alpha^2$ (\ref{fig:fock darwin spectrum}b). (iii) Finally, at some magnetic field the level crossing
of the Fock-Darwin levels is lifted by spin-orbit coupling. The resulting level repulsion
is linear in $\alpha$ (\ref{fig:fock darwin spectrum}c). The states here are the spin hot spots, that is states in which both Pauli spin up and down species contribute significantly.\cite{fabian1998:PRL,bulaev2005:PRB,destefani2004:PRB}

The above picture can be understood from general symmetry considerations within the framework of
perturbation theory.  All spin-orbit terms commute, at $B=0$, with the time inversion operator $T=i \sigma_y \hat{C}$, where $\hat{C}$ is the operator of complex conjugation. Therefore Kramer's degeneracy is preserved so that
the states are always doubly degenerate. The linear terms can be transformed 
into each other by a unitary transformation $(\sigma_x + \sigma_y)/\sqrt{2}$ (spin rotation around $[110]$ by $\pi$ ), which commutes with $H_0$. Therefore the effects on the energy spectrum induced individually by the linear Dresselhaus and the Bychkov-Rashba terms are identical at $B=0$. At finite magnetic fields the two interactions give qualitatively
different effects in the spectrum, especially for spin hot spots, as discussed below. 

For any $B$ the following commutation relations hold for the linear terms:
\begin{equation} \label{eq:commutators} 
[H_{BR},l_z+s_z]=0,\quad [H_D,l_z-s_z]=0.
\end{equation}
This means that $H_{BR}$ commutes with the angular momentum, while $H_D$ does not.
 This will influence  the interference between the two terms in the energy spectrum.
We can use the Fock-Darwin states as a basis for perturbation theory. Up to the second order 
the energy of state $|i\rangle=\Psi_{n,l,\sigma}$ is
\begin{equation} \label{eq:non degenerate perturbation}
E_i=\epsilon_i+\langle i |H_{SO}|i\rangle + \sum_{j\neq i} \frac{\langle i
|H_{SO}|j\rangle\langle j|H_{SO}|i\rangle}{\epsilon_i-\epsilon_j}.
\end{equation}
The first order correction is zero for all spin-orbit terms since $H_{SO}$ contain only odd 
powers of $\textbf{K}$ whose expectation values in the Fock-Darwin states vanish. If the perturbation
expansion is appropriate, the spin-orbit interactions have a second order (in $\alpha$)
effect on energy.

Both linear spin-orbit terms couple states with orbital momenta $l$ differing by one. 
It then follows from the commutation relations \eqref{eq:commutators} that $H_{BR}$ preserves the total
angular momentum $l+s$, while $H_D$ preserves the quantity $l-s$. As a result, there is no correction
to the energy from the interference terms between $H_{BR}$ and $H_{D}$ in 
Eq. \eqref{eq:non degenerate perturbation}: $\langle i | H_{BR} | j \rangle  \langle j | H_D | i \rangle = 0$.  
As for the cubic Dresselhaus term, only the following orbital states are coupled:
$(l,\uparrow)\rightarrow\{(l+3,\downarrow),(l-1,\downarrow)\}$ and
$(l,\downarrow)\rightarrow\{(l-3,\uparrow),(l+1,\uparrow)\}$. Due to these selection rules there 
are no interference terms $\sim H_{D3}H_{BR}$, but terms $\sim H_{D3}H_D$ will contribute
to energy perturbation. The Bychkov-Rashba and Dresselhaus Hamiltonians act independently on the 
Fock-Darwin spectrum (up to the second order). 

To gain more insight into the perturbed structure of the spectrum at $B=0$, we rewrite 
Eq. \eqref{eq:non degenerate perturbation} using an auxiliary anti-hermitian operator $H_{SO}^{op}$ defined
by the commutation relation $[H_0,H^{op}_{SO}]=H_{SO}$.
If such an operator exists, the second order correction in \eqref{eq:non degenerate perturbation} is then 
\begin{eqnarray}
&&\sum_{j\notin\mathcal{N}}\frac{\langle i|H_{SO}|j\rangle\langle j|H_{SO}|i\rangle}{\epsilon_i-\epsilon_j}
=\langle i|\frac{1}{2}[H^{op}_{SO},H_{SO}] |i \rangle+
\nonumber\\ \label{eq:sum rule}
&&\qquad+Re(\langle i |H_{SO}P_\mathcal{N} H^{op}_{SO}|i\rangle),
\end{eqnarray}
where $P_\mathcal{N}$ is the projector on the subspace $\mathcal{N}$ of the states excluded
from the summation. In our case here it is just one state, $\mathcal{N}=\{|i\rangle\}$.
The last term  in \eqref{eq:sum rule} then vanishes. The auxiliary operator for $H_{D3}$ is not known and if found, it must depend on the confining potential. Operators for the linear
terms are:\cite{aleiner2001:PRL}
\begin{eqnarray} \label{eq:operators op}
H^{op}_D & = & -i(\alpha_D/2l_0)(x\sigma_x-y\sigma_y), \\
\label{eq:operators2}
H^{op}_{BR} & = & i(\alpha_{BR}/2l_0)(y\sigma_x-x\sigma_y).
\end{eqnarray}
The corresponding commutators are (in the zero magnetic field $\textbf{K}=\textbf{k},\,L_z=l_z,\,\theta=0$; the last expression will be useful later)
\begin{eqnarray} \label{eq:commutators with op}
\left [ H_D^{op} , H_D\right  ] & = & - E_0 \alpha_D^2 (1-\sigma_z L_z), \\
\left [ H_{BR}^{op} , H_{BR}\right  ] & = & -  E_0 \alpha_{BR}^2 (1+\sigma_z L_z),\\
\left [ H_D^{op} , H_{D3} \right] &=&E_0 l_0^2\alpha_D\alpha_{D3}\Big(K_x^2+K_y^2-\nonumber\\
&&\hspace{-2cm}-2\sigma_z[x K_y K_x^2-y K_x K_y^2-2i\theta(x K_x+y K_y)]\Big).
\label{eq:dd3 commutator}
\end{eqnarray} 
Because $[H_D^{op},H_{BR}] + [H_{BR}^{op},H_D]=0$, the corrections to the second order perturbation add independently for $H_{BR}$ and $H_D$ (as also noted above from the selection rules), we can introduce $H_{\mathrm{lin}}^{op}=H_D^{op}+H_{BR}^{op}$.  
An alternative route to Eq. \eqref{eq:sum rule} is to transform the Hamiltonian with\cite{aleiner2001:PRL}
$U=\exp(-H_{SO}^{op})$ to $\tilde{H} = H_0 - (1/2)[H_{SO},H_{SO}^{op}]$ in 
the second order of $\alpha$. The final result can be also obtained in a straightforward way by using the 
Thomas-Reiche-Kuhn sum rule in the second order of perturbation theory with the original spin-orbit terms.
The resulting effective Hamiltonian is (terms depending on $\alpha_{D3}$ are omitted here)
\begin{equation} \label{eq:effective}
\tilde{H}=H_0-E_0(\alpha_D^2+\alpha_{BR}^2)/2+E_0\sigma_z
L_z(\alpha_D^2-\alpha_{BR}^2)/2.
\end{equation}
This Hamiltonian, in which the spin-orbit coupling appears in its standard form, neatly explains point (ii)
about the lifting of the degeneracy at $B=0$. The levels in Fig. \ref{fig:fock darwin spectrum}b, for example, are four
fold degenerate ($|l|=1$, $|\sigma|=1$) without spin-orbit coupling. Turning on, say, $H_D$, will split the
levels into two groups: energy of the states with $l\sigma>0$ would not change in the second order, while
the states with $l\sigma<0$ will go down in energy by $E_0\alpha_D^2$, as seen in Fig.  \ref{fig:fock darwin spectrum}b. 

\subsection{Spin hot spots}

\begin{figure}
\centerline{\psfig{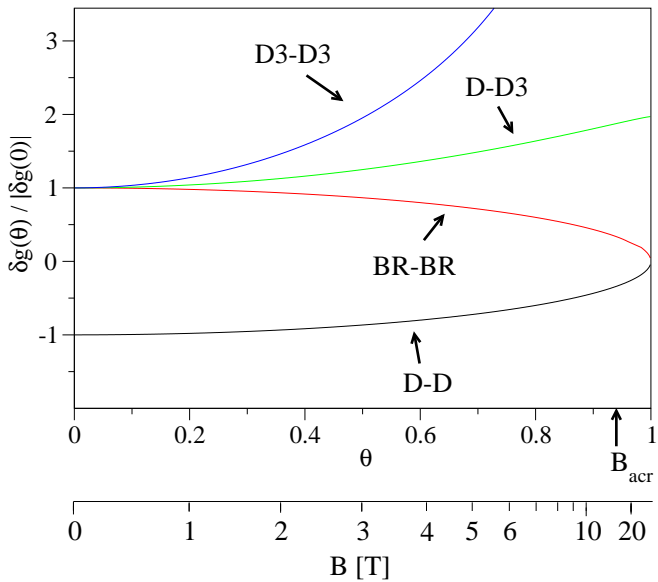}}
\caption{Calculated corrections to the effective g-factor by spin-orbit interactions.
Formulas \eqref{eq:g factor corrections} scaled by the values at $B=0$ (and thus independent on
$\alpha_{SO}$) are plotted. The actual numerical values of $\delta g$ at $B=0$ are: $\delta
g_{D-D}(0)=1.0\times 10^{-2}$, $\delta g_{BR-BR}(0)=8.6\times 10^{-4}$, $\delta
g_{D-D3}(0)=9.4\times 10^{-4}$, $\delta g_{D3-D3}(0)=2.5\times 10^{-5}$. At the anti-crossing $\delta g_{D-D}(B_\mathrm{acr})=2.4\times 10^{-3}$, $\delta g_{BR-BR}(B_\mathrm{acr})=1.0\times 10^{-4}$, $\delta g_{D-D3}(B_\mathrm{acr})=1.8\times 10^{-3}$, $\delta g_{D3-D3}(B_\mathrm{acr})=3.4\times 10^{-4}$.}
\label{fig:g factor corrections} \end{figure}

Spin hot spots are states formed by two or more states whose energies in the absence of spin-orbit
coupling are degenerate or close to being degenerate, while turning on the coupling removes the
degeneracy.\cite{fabian1998:PRL} Such states are of great importance for spin relaxation, which
is strongly enhanced by their presence.\cite{fabian1999:PRL, bulaev2005:PRB} The reason is that the degeneracy
lifting mixes spin up and spin down states and so transitions between states of opposite magnetic
moment will involve spin flips with a much enhanced probability compared to normal states.  

Figure \ref{fig:fock darwin spectrum}c shows an interesting situation where two degenerate levels are lifted by 
spin-orbit coupling.\cite{destefani2004:PRB, bulaev2005:PRB} The 
lifting is of the first order in $\alpha$, unlike the lifting of degeneracy
at $B=0$ in which case the degenerate states are not directly coupled by $H_{SO}$. In a finite magnetic
field, at a certain value $B_\mathrm{acr}$, the states of opposite spins and orbital momenta differing
by one cross each other, as follows from the equation \eqref{eq:fock darwin  energies}. The crossing field
is $B_\mathrm{acr}\approx|\alpha_Z|^{-1/2}\hbar/(e l_0^2)$, which is about 13.4 T for our parameters (making 
the confinement length larger the magnitude of the field would decrease). Spin-orbit interaction couples
the two states thereby lifting the degeneracy. For GaAs, where $g^* < 0$, only the Bychkov-Rashba
term couples the two states. The Dresselhaus terms are not effective ($H_{D3}$ would introduce such 
a splitting at $3B_\mathrm{acr}$). The energy splitting due to $H_{BR}$ is 
\begin{equation}
\Delta = c E_0 \alpha_{BR}(1-\theta_\mathrm{acr})(1-\theta_\mathrm{acr}^2)^{1/4},
\end{equation}
where $c$, which is a number of order 1, depends on the quantum numbers of the two states. 
Since spin hot spots at $B_\mathrm{acr}$ are due only $H_{BR}$, the splittings could help to sort out
the Bychkov-Rashba versus Dresselhaus contributions. Figure \ref{fig:fock darwin spectrum}c shows the calculated
level repulsion for states $n=0, l=0,\sigma=\downarrow$ and $n=0, l=-1,\sigma=\uparrow$. The magnitude of $\Delta$, though
being linear in $\alpha_{BR}$, is on the order of $10^{-3}$ meV and thus comparable to the energy
scales associated with quadratic spin-orbit perturbations.

\subsection{Effective g-factor}

When probing spin states in quantum dots with magnetic field, important information comes from the measured
Zeeman splitting. We will focus here on the two lowest spin states and calculate the effective $g$-factor
as $g=(E_{0,0,\downarrow}-E_{0,0,\uparrow})/(\mu_B B)$. If $H_{SO}=0$, then the effective $g$-factor equals to the conduction band value $g^*$. The actual value in the presence of spin-orbit coupling is important for understanding single spin 
precession in magnetic field, which seems necessary to perform single qubit operations in quantum dots.  
We have obtained the following contributions to the $g$-factor from non-degenerate (that is, excluding spin 
hot spots) second-order perturbation theory [Eq. \eqref{eq:non degenerate perturbation}] (for linear spin-orbit terms these are derived also in\cite{sousa2003:PRB,puente2002:PRB}):
\begin{eqnarray}
\delta g_{D-D}&=&-\frac{\alpha_D^2}{2 m/m_e}\frac{\sqrt{1-\theta^2} [1-\theta^2-2(1+\theta^2)\alpha_Z]}{1-\theta^2(1+4\alpha_Z+4\alpha_Z^2)},\nonumber\\
\delta g_{BR-BR}&=&\frac{\alpha_{BR}^2}{2m/m_e} \frac{\sqrt{1-\theta^2}[1-\theta^2+2(1+\theta^2)\alpha_Z]} {1-\theta^2(1-4\alpha_Z+4\alpha_Z^2)},\nonumber\\
\delta g_{D-D3}&=&\frac{\alpha_{D}\alpha_{D3}}{m/m_e} \frac{(1+\theta^2)[1-\theta^2-2(1+\theta^2)\alpha_Z]}{1-
\theta^2(1+4\alpha_Z+4\alpha_Z^2)},\nonumber\\
\delta g_{D3-D3}&=&\frac{\alpha_{D3}^2} {8m/m_e\theta\sqrt{1-\theta^2}} \bigg(\frac{2(1-\theta)^2(1+\theta^2)^2}{1-\theta(1+2\alpha_Z)}+\nonumber\\
&+&\frac{(1-\theta)^4(1+\theta)^2}{3-\theta(1+2\alpha_Z)}+
\frac{-3(1-\theta)^6}{3-\theta(3-2\alpha_Z)}+\nonumber\\
&+&\frac{3(1+\theta)^6}{3+\theta(3-2\alpha_Z)}-
-\frac{2(1+\theta)^2(1+\theta^2)^2}{1+\theta(1+2\alpha_Z)}-\nonumber\\
&-&\frac{(1-\theta)^2(1+\theta)^4}{3+\theta(1+2\alpha_Z)}\bigg).
\label{eq:g factor corrections}
\end{eqnarray} 
Here $\delta g_{A-B}$ stands for a correction that is proportional to $\alpha_A \alpha_B$.

The functions \eqref{eq:g factor corrections} are plotted in Fig. \ref{fig:g factor corrections}.
We can understand the limits of $\delta g$ at $B\to\infty\,(\theta\to 1)$ if we notice that in the natural length unit $l_B$ the momentum $K_x=-i\partial_x-yBe/2\hbar=l_B^{-1}[-i\partial_{x/l_B}-\theta (y/l_B)]$. In the limit $B\to\infty$ the matrix elements of
$H_D$, which is linear in $K$, scale as $l_B^{-1}$, while the Fock-Darwin energies scale as $l_B^{-2}$.
The second order $D$-$D$ correction to $E_{0,0,\downarrow}-E_{0,0,\uparrow}$ is thus independent of $l_B$;
it converges to  $-E_0\alpha_D^2/(1+\alpha_Z)$. The $BR$-$BR$ correction is analogous, with the limit $E_0\alpha_{BR}^2/(1-\alpha_Z)$. To get the g-factor we divide the energy differences by $\mu_B B$ and
get $\delta g_{D-D}\,(\theta\to 1)\propto B^{-1}$; similarly for $H_{BR}$. Since $H_{D3}$ scales as
$l_B^{-3}$ one gets $\delta g_{D-D3}\,(\theta\to 1)\to 2\alpha_D\alpha_{D3}m/(1+\alpha_Z)m_e$ and
$\delta g_{D3-D3}(\theta\to 1)\propto B$. It seems that for increasing
$B$ there inevitably comes a point where the influence of $H_{D3}$ on the g-factor dominates. But at $B=B_\mathrm{acr}$ there is an anti-crossing of the states $(0,0,\downarrow)$ and $(0,-1,\uparrow)$ so for larger $B$ the 
g-factor does not describe the energy difference between the two lowest states, but between the second excited
state and the ground state. 
The value of $B$ where  $\delta g_{D3-D3}=\delta g_{D-D}$ is given by $B\approx (\hbar/e l_0^2) \alpha_D/\alpha_{D3}\sqrt{2}$. For GaAs parameters it is $\approx 25$ T. After anti-crossing the first exited state is $\Psi_{0,-1,\uparrow}$ and its energy difference to the ground state is divided by $\mu_B B$ is $\propto 1/B^2$ for $H_0$. The corrections from spin-orbit terms are $D$-$D,\,BR$-$BR\propto 1/B^5$, $D$-$D3\propto 1/B^4$, and 
$D3$-$D3 \propto 1/B^3$.

\section{Double dots}

A double dot structure comprises two single dots close enough for their
mutual interaction to play an important role. Here we consider symmetric dots modeled by 
$V_C^{dd}$ of Eq. \eqref{eq:doubledot}. Such a potential has an advantage that in the
limits of small $d\to 0$ and large $d\to\infty$, the solutions 
converge to the single dot solutions centered at $x=0$ and $\pm l_0 d$, respectively. 
We denote the displaced Fock-Darwin states as
$\Psi_{n,l,\sigma}^{\pm d}(x,y)\equiv\Psi_{n,l,\sigma}(x\pm l_0 d,y)$.

\begin{table}
\begin{tabular}{|c|c|c|}
\hline
magnetic field&SO terms&symmetries of $H$\\
\hline
$B=0$&none&$I_x$,$I_y$,$I$,$T$,$R_{\textbf{n}}$\\
&$BR$&$-i\sigma_x I_x$,$-i \sigma_y I_y$,$-i\sigma_z I$,$T$\\
&$D,D3$&$-i\sigma_y I_x$,$-i \sigma_x I_y$,$-i\sigma_z I$,$T$\\
&all&$-i\sigma_z I,T$\\
\hline
$B>0$&none&$-i\sigma_z I$,$R_z$\\
&any&$-i\sigma_z I$\\
\hline
\end{tabular}
\caption{Symmetries of the double dot Hamiltonian for different 
spin-orbit terms present at $B=0$ and $B>0$. 
Here $I_x(I_y)$ means reflection $x\to-x$ ($y\to-y$), $I=I_x I_y$, and $R_z=\exp(-i\phi \sigma_z/2)$ is the
rotation of a spinor by angle $\phi$ around z-axis; $R_{\textbf{n}}$ is a spinor rotation around an 
arbitrary axis $\textbf{n}$ and $T$ is the time reversal symmetry. The identity operation is not listed. }
\label{tab:symmetries}
\end{table}

The symmetries of the double dot Hamiltonian with spin-orbit couplings are listed in Tab. \ref{tab:symmetries}.
The time symmetry is always present at $B=0$, giving Kramer's double degeneracy. 
The rotational space symmetry from the single dot case is lost; instead there are two discrete symmetries -- reflections $I_x$ about $y$ and $I_y$ about $x$.
In zero magnetic field and without spin-orbit terms, the Hamiltonian has both $I_x$ and $I_y$ symmetries.
If only Rashba or Dresselhaus terms are present, we can still preserve symmetries $I_x$ and $I_y$ by 
properly defining the symmetry operators to act also on the spinors (forming the double group). 
The Bychkov-Rashba term, $H_0+H_{BR}$, is invariant to operations defined by the spatial invariance.
This is not the case of $H_D$, since here the operators $-i\sigma_y I_x$ and $-i\sigma_x I_y$ do not
describe a spatial reflection of both the orbital and spinor parts. The symmetry operations for 
$H_{BR}$ and $H_D$ are connected by the unitary transformation $(\sigma_x+\sigma_y)/\sqrt{2}$,
which connects the two Hamiltonians themselves. 
Finally, if both spin-orbit terms are present, or at $B>0$, the only space symmetry left is $I=I_x I_y$.

In the following we consider separately the cases of zero and finite magnetic fields.

\subsection{Energy spectrum in zero magnetic field, without spin-orbit terms.}

If no spin-orbit terms are present the group of our double dot Hamiltonian is $C_{2v}\otimes SU(2)$. 
The $SU(2)$ part accounts for the (double) spin degeneracy.
The orbital parts of the eigenstates of the Hamiltonian therefore transform according to the irreducible representations of $C_{2v}$. The representations\cite{koster1957} $\Gamma_i$, $i=1...4$, along with
their transformation properties under the symmetries of $C_{2v}$, 
are listed in Tab. \ref{tab:representations}. The symmetry properties will be used in discussing 
the perturbed spectrum.

\begin{table}
\begin{tabular}{|c|c|cc|cc|}
\hline
repre-&under $I_x$, $I_y$&\multicolumn{4}{c|}{numbers for $g_i^{n,l,\sigma}$}\\ \cline{3-6}
sentation&transforms&$l$ - even&&$l$ - odd&
\\\cline{3-6}
&as&L&D&L&D\\\hline
$\Gamma_1$ &1&1&1&-1&-1\\
$\Gamma_2$ &x&-1&-1&1&1\\
$\Gamma_3$ &xy&-1&1&1&-1\\
$\Gamma_4$ &y&1&-1&-1&1\\\hline
\end{tabular}
\caption{Notation and transformation properties of $C_{2v}$ representations. $L$ and $D$ are the 
coefficients of the dependence of $g_i^{n,l,\sigma}$ on the single dot functions (see text).}
\label{tab:representations}
\end{table}

We denote the exact eigenfunctions of the double dot Hamiltonian with 
$H_{SO}=0$ as $\Gamma_{i\sigma}^{ab}$ where $a(b)$ is the single dot
level to which this eigenfunction converges as $d\to 0\,(\infty)$; $i$ labels
the irreducible representation, $\sigma$ denotes spin.  
We have chosen the confining potential to be such, that at $d\to 0(\infty)$ the solutions of the double dot $H_0$ converge to
the (shifted) Fock-Darwin functions, if properly symmetrized according to the representations of $C_{2v}$.
These symmetrized functions will be denoted as $g^{n,l,\sigma}_i$, where (up to normalization)
\begin{equation} \label{eq:gnl}
g^{n,l,\sigma}_i=(\Psi_{n,l,\sigma}^d+D_i\Psi_{n,l,\sigma}^{-d})+L_i(\Psi_{n,-l,\sigma}^d+D_i\Psi_{n,-l,\sigma}^{-d}).
\end{equation}
 The numbers $D_i(L_i)$ for different irreducible representations are in the Tab. \ref{tab:representations}.
 
Generally, up to normalization, the exact solution can be written as a linear combination of any complete set of functions (we omit the spin index which is the same for all terms in the equation) 
\begin{equation} \label{eq:LCSDO}
\Gamma^{ab}_i=\sum_{n,l}\widetilde{c}(n,l)g_i^{n,l}=g_i^{n_0,l_0}+\sum_{n,l}c(n,l)g_i^{n,l}.
\end{equation}
The last equation indicates the fact, that for a function $\Gamma_i^{ab}$ in the limit $d\to0(\infty)$, there will be a dominant $g$-function in the sum with the numbers $n_0,l_0$ given by the level $a(b)$ and the coefficients $c$ for the other functions will converge to zero. We term the approximation $c(n,l)=0$ as a linear combination of single dot orbitals (LCSDO).
  
Knowing the representations of the double dot Hamiltonian and the fact that
Fock-Darwin functions form $SO(2)$ representations (reflecting the symmetry of single dot $H_0$)
we can decompose all single dot levels into the double dot representations and thus
formally construct the energy spectrum of a double dot using the symmetry considerations only.

\begin{figure}
\centerline{\psfig{file=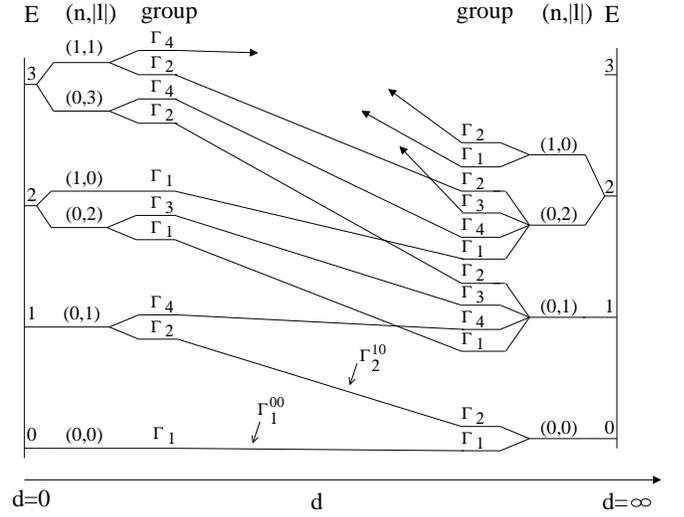,width=1\linewidth}}
\caption{Single electron spectrum of a symmetric ($C_{2v}$) lateral double dot structure as a function of the
interdot separation, at $B=0$, derived by applying group theoretical considerations. Single dot states at
$d=0$ and $d=\infty$ are labeled by the principal ($n$) and orbital ($l$) quantum numbers, while the
double dot states are labeled according to the four irreducible representations $\Gamma_i$ of $C_{2v}$. The
lowest double dot states have explicitly written indices showing the excitation level of the $d=0$ and $d=\infty$
states they connect. Every state is doubly (spin) degenerate, and spin index is not given. }
\label{fig:schematic spectrum}\end{figure}

Following the standard technique for constructing such correlation diagrams (connecting 
states of the same representation and avoiding crossing of lines of the same representation)
we arrive at the spectrum shown in Fig. \ref{fig:schematic spectrum}. The ground state transforms
by the symmetry operations according to $\Gamma_1$ (identity), while the first excited state
according to $\Gamma_2$ ($x$). This is expected for the symmetric and antisymmetric states formed
by single dot ground states. The symmetry structure of the higher excited states is important
to understand spin-orbit coupling effects. Indeed, the spin-orbit terms couple two opposite spins
according to certain selection rules. Since $H_D$, for example, transforms similarly to $x\oplus y$,
it couples the ground state $\Gamma_1$ with $\Gamma_2$ and $\Gamma_4$. In general, odd numbered representations
can couple to even numbered representations. The same holds for $H_{BR}$ and $H_{D3}$. If we include either $H_{BR}$
or $H_D$ into the Hamiltonian, and consider spinors as the basis for a representation, the  states would transform according 
to $\Gamma_5$, the only irreducible representation of the double group of $C_{2v}$.

\begin{figure}
\centerline{\psfig{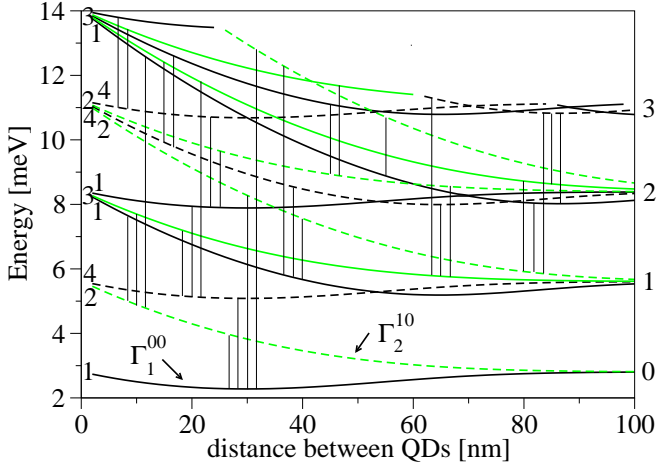}}
\caption{Calculated energy spectrum of a double quantum dot at $B=0$, as a function of interdot distance. Spin-dependent terms are not included. Vertical bars indicate couplings due to spin-orbit coupling. Group theoretical symbols are shown with the lines on the left. Single dot levels are denoted by the highest orbital momentum (0, 1, 2, ...) present in the degenerate set. This labeling is on the right. Every state is doubly degenerate, and spin index is not given.}
\label{fig:spectrum}
\end{figure}

The calculated numerical spectrum for our model structure is shown in Fig. \ref{fig:spectrum}. There is 
a nice qualitative correspondence with Fig. \ref{fig:schematic spectrum}.
In Fig. \ref{fig:spectrum} by vertical bars we denote coupling through $H_D$  or $H_{BR}$ ($|\langle i| H_D |j\rangle|= |\langle i| H_{BR} |j\rangle|$). The couplings follow the selection rule described above.
Since there are several level crossings in the lowest part of the spectrum, a question arises if spin hot 
spots are formed in the presence of spin-orbit coupling. It turns out, that there is no first-order
level repulsion at the crossings due to the linear spin-orbit terms because the levels are not coupled by the linear terms, even though 
such couplings are allowed by symmetry. \textit{There are no spin hot spots due to the linear spin-orbit terms at zero magnetic field.} For example $\Gamma_4^{11}$ and $\Gamma_1^{21}$ are not coupled by spin-orbit terms, and therefore their degeneracy (at $2dl_0 \approx 50$ nm) is not lifted by linear spin-orbit terms as we would expect from symmetry (actually, there is an anti-crossing which is of the order $\alpha_\mathrm{lin}^3$, instead of the expected $\alpha_\mathrm{lin}$). The cubic Dresselhaus term gives here (and also in other crossings that conform with the selection rules) a linear anti-crossing, as one expects. The absence of anti-crossings from the linear spin-orbit terms will be explained in the next section.

Since our main goal here is to study the effects of spin-orbit coupling on the tunneling
between the two dots, we first look at the tunneling for $H_{SO}=0$. The tunneling energy, $\delta E_t/2$,
is given by the difference between the energies of the symmetric ground state and the asymmetric first excited 
orbital state: $\delta E_t = E_A - E_S$. We compare the LCSDO approximation with our numerically exact calculations. The functions Eq. \eqref{eq:gnl} are not orthogonal. If we introduce the overlap integrals between 
these functions (all the indices of a $g$-function are denoted by one letter here) by $S_{ij} = \langle i|j\rangle$, and the Hamiltonian matrix elements 
$H_{ij} = \langle i |H|j\rangle$, variational theory gives for the expansion coefficients $c_i$
of a double dot state $\Gamma = \sum_i c_i |i\rangle$ the generalized eigenvalue equation, 
\begin{equation} \label{eq:generalized eigenvalue}
\sum_j H_{ij} c_j = E \sum_j S_{ij} c_j.
\end{equation}
  
We will compute the energy of the two lowest orbital double dot states, $\Gamma_{1\sigma}^{00}\equiv \Gamma_S^{\sigma}$, $\Gamma_{2\sigma}^{10}\equiv\Gamma_A^{\sigma}$, which are a 
symmetric and an antisymmetric state with respect to $x$. The energies are denoted as  $E_S^{(0)},E_A^{(0)}$, where index zero indicates the absence of spin-orbit
coupling.

We first use the LCSDO approximation suitable for the limit $d\to\infty$, that is we approximate $\Gamma_S^\sigma\approx g_1^{0,0,\sigma}$, and $\Gamma_A^\sigma\approx g_2^{0,0,\sigma}$. This means the basis in Eq. \eqref{eq:generalized eigenvalue} consists of one function and the solution for the energy is $E_i=H_{ii}/S_{ii}$. For the considered states we obtain:
\begin{eqnarray}  
E_S^{(0)}&=&2E_0\frac{1+[1-2d/\sqrt{\pi}]e^{-d^2}+d^2 \mathrm{Erfc}(d)}{1+e^{-d^2}},
\nonumber\\\label{eq:full expressions no B}
E_A^{(0)}&=&2E_0\frac{1-e^{-d^2}+d^2 \mathrm{Erfc}(d)}{1-e^{-d^2}}.
\end{eqnarray}
In the limit of large interdot separation (that is we make expansion in powers of $e^{-d^2}$ and take the first term in this expansion as the leading order), $E_{S(A)}^{(0)} \approx E_0[2\pm (2/\sqrt{\pi}) d\exp(-d^2)]$,
where the minus (plus) sign is for $E_{S}^{(0)}$ ($E_A^{(0)}$). The tunneling energy then is
\begin{equation} \label{eq:dEt}
\delta E_t^{(0)} \approx E_0 \frac{4}{\sqrt{\pi}} d e^{-d^2}.
\end{equation}
To understand this result, we get it once again by taking a two dimensional basis in Eq. \eqref{eq:generalized eigenvalue} consisting of functions $\Psi_{0,0,\sigma}^d$ and 
$\Psi_{0,0,\sigma}^{-d}$.  In the limit $d\to\infty$ both diagonal  elements of the matrix $H$ converge to the energy of the Fock-Darwin ground state ($2E_0$). Then the difference of the eigenvalues of $H$ is given by the off-diagonal matrix element, which is $H_{12}=\langle \Psi_{0,0,\sigma}^{-d}|H \Psi_{0,0,\sigma}^d\rangle \sim (\Psi_{0,0,\sigma}^{-d}\Psi_{0,0,\sigma}^d)(\vec{r}=0)\sim e^{-d^2}$.

Any further refinement beyond LCSDO would not contribute significantly to the calculated
$\delta E_t^{(0)}$. For example, for the ground state to go beyond LCSDO, we take the next excited function of the same symmetry and get the basis for Eq. \eqref{eq:generalized eigenvalue} to be $\{g_1^{0,0,\sigma},g_1^{0,1,\sigma}\}$. The off-diagonal element in the matrix $H$ is $H_{12}=\langle
g_1^{00}|H_0|g_1^{01}\rangle\approx E_0 P(d)e^{-d^2}$, where $P(d)$ is a polynomial in $d$.
 Since now $H_{11}-H_{22}$ is
of the order of $E_0$, the correction from non-diagonal terms will be of the order of
$(e^{-d^2})^2$ and will not change the leading order result.

\begin{figure}
\centerline{\psfig{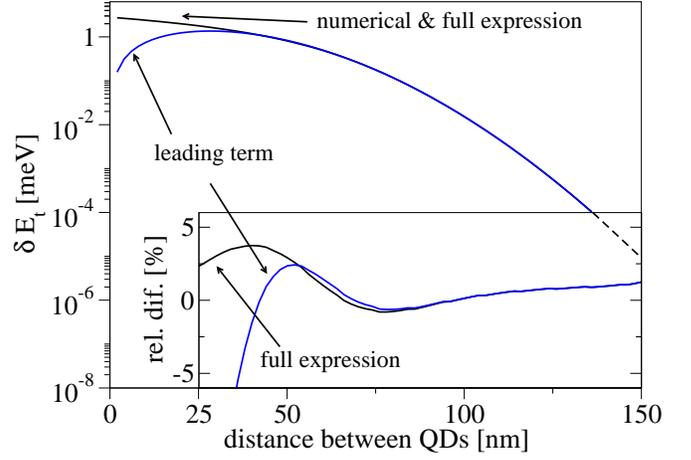}}
\caption{Calculated tunneling energy  $\delta E_t^{(0)}$ for the double dot Hamiltonian $H_0$.
Numerical results are compared with the analytical (full) expression for $\delta E_t^{(0)}=E_A^{(0)}-E_S^{(0)}$, where
the energies of the asymmetric and symmetric states are given by Eq. \eqref{eq:full expressions no B}.
The leading order expression is that of Eq. \eqref{eq:dEt}, valid for large $d$. 
The inset shows the relative error (with respect to the numerical calculation) of the two analytical results.}
\label{fig:tunneling}
\end{figure}

The numerical result, together with the analytical forms of $\delta E_t^{(0)}$ [one using the complete
expressions Eq. \eqref{eq:full expressions no B}, the other for the limiting case of large $d$,
Eq. \eqref{eq:dEt}], is plotted in Fig \ref{fig:tunneling}. The complete expression is in excellent
agreement with the numerics, over the whole range of $d$, including the limit $d\to 0$.
As for the leading order
expression, it becomes an excellent description for the tunneling energy at distances roughly
twice the dot confinement length (40 nm in our case), as seen from the inset to Fig. \ref{fig:tunneling}.

Using\label{uncorresponding LCSDO} a LCSDO approximation $g_1^{0,0,\sigma}$ for the ground state is correct for both limits $d\to 0$ and $d\to \infty$, because $\Gamma_{1\sigma}^{00}$ converges to the single dot level 0 in both limits. Therefore it is reasonable to expect that the approximation will be equally good for the whole range of $d$. However, for the first excited state $\Gamma_2^{10}$ the two limits go into different single dot levels and the proper LCSDO approximations for this state are $g_2^{0,0,\sigma}$ and $g_2^{0,1,\sigma}$ in the limit $d\to\infty$ and $d\to0$ respectively. However  $g_2^{0,0,\sigma}\sim g_2^{0,1,\sigma}+g_4^{0,1,\sigma}$ as $d\to0$ and thus for both ground and first excited states using a LCSDO approximation suitable for $d\to \infty$ gives good results in the whole range of $d$. We will see, that this will not be true for a non-zero magnetic field and describing $\Gamma_{2\sigma}^{10}$ by $g_2^{0,0,\sigma}$ will give a much higher error. A remedy is to go beyond LCSDO for $\Gamma_2^{10}$, for example by taking the base for Eq. \eqref{eq:generalized eigenvalue} to include two $g$-functions, each correct in one of the limits $d\to 0$, $d\to\infty$. We will not present the results of such computations since formulas become more complicated without giving better understanding.
 
Higher orbital states can be treated similarly. Starting from level 2 there are more
functions of the same representation in one single dot level, therefore the basis for the Hamiltonian Eq. \eqref{eq:generalized eigenvalue}
giving the leading order must contain more that one function. 

\subsection{Corrections to energy from spin-orbit coupling in zero magnetic field. \label{sec:DDSOC1}}

When we add $H_{SO}$ to $H_0$, the structure of the corrections to the energies of the two lowest states 
up to the second order in spin-orbit couplings can be expressed as
\begin{equation}
E^{(2)}_i=-\mathcal{A}_i(\alpha_D^2+\alpha_{BR}^2)-\mathcal{B}_i \alpha_{D3}^2+\mathcal{C}_i
\alpha_D \alpha_{D3},\label{eq:structure of SO corrections}
\end{equation}
where $i$ is either $S$ or $A$. The coefficients $\mathcal{A},\mathcal{B},$ and $\mathcal{C}$ are positive for all values of the interdot distance. The differences $\mathcal{A}_A-\mathcal{A}_S,\ldots$ approach zero as $d\to \infty$. We will
argue below that $\mathcal{A}_S=\mathcal{A}_A=1/2$ with the exception of a very small interdot distance (less than 1 nm). \textit{There are thus no contributions from the linear
spin-orbit couplings to $\delta E_t$ in the second order.} Only the cubic Dresselhaus term contributes,
either by itself or in combination with the linear Dresselhaus term. Spin-dependent tunneling is 
greatly inhibited.

\begin{figure}
\centerline{\psfig{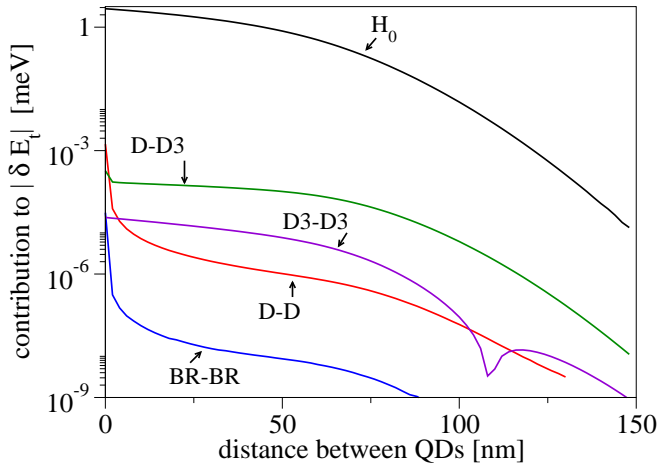}}
\caption{Calculated corrections to the tunneling energy $\delta E_t$ from spin-orbit terms at $B=0$.
 The labels indicate which spin-orbit terms are involved. Only $D$-$D3$ and $D3$-$D3$ are of second order. The remaining contributions are of fourth order.}
\label{fig:tunneling corrections}\end{figure}

Numerical calculation of the corrections to $\delta E_t$ from spin-orbit couplings 
 are shown in Fig \ref{fig:tunneling corrections}.
The dominant correction is the mixed $D$-$D3$ term, followed by $D3$-$D3$. These are the only second 
order corrections. For GaAs, and our model geometry, these corrections are about 4 and 5 orders of magnitude lower than $\delta E_t^{(0)}$. The corrections, when only linear spin-orbit terms are present, are much smaller since they are of the fourth order.
The dramatic enhancement of the corrections from linear spin-orbit terms close to $d=0$ is due to the
transition from coupled to single dots. We will explore this region in more detail later.

We will first show how naive approaches to calculating spin-orbit contributions to tunneling fail to
explain the above results. We use the example of the linear Dresselhaus term. The simplest way to include
this term is to begin with the two lowest orbital states (that are four states including spin),
$g_1^{0,0,\sigma}$ and $g_2^{0,0,\sigma}$.  Because of
the time reversal symmetry the resulting $4\times 4$ matrices $H$ and $S$ from Eq.\eqref{eq:generalized eigenvalue} block diagonalize
into two equal $2\times 2$ matrices with elements
$H_{11}=E_S^{(0)}$, $H_{22}=E_A^{(0)}$, and $H_{12}=\langle
g_{1}^{0,0,\uparrow}|H_D|g_{2}^{0,0,\downarrow}\rangle=-i E_0\alpha_D d e^{-d^2}$; 
$S$ is the unit matrix now, since the two states are orthogonal due to symmetry.
Using the large $d$ limit for $\delta E_t^{(0)}$, Eq. \eqref{eq:dEt}, we obtain the perturbed
energies $E_{S(A)}=2E_0 \pm E_0\sqrt{4/\pi+\alpha_D^2}de^{-d^2}$ with the minus (plus)
sign for $S$ ($A$). In the second order of $\alpha_D$ the symmetric and antisymmetric
level energies have opposite contributions, giving $\delta E_t \approx [(4E_0/\sqrt{\pi})
+ (E_0\sqrt{\pi}/2) \alpha_D^2 ]d\exp(-d^2)$, in contrast to the numerical results where there is
no dependence on $\alpha_D^2$ in the second order.   
Enlarging the basis by the first excited orbital states (all together 12 states including spin),
that is, include $g_i^{0, 1,\sigma}$, the symmetric and antisymmetric level energies
will have the correct limit for the spin-orbit contributions, $-E_0\alpha_D^2/2$, at $d\to\infty$. 
At finite values of $d$ the difference from this limit value is less than 2\%. That means there are still terms of order $\alpha_D^2$ in $\delta E_t$.
Could using a renormalized basis help? We could, for example, use symmetrized 
states of the separated dots that include spin-orbit terms. It is not difficult
to see that this would not work either: the perturbed single dot ground state, for example,
contains the spin admixture from the first excited orbital states. This is then 
similar to using the 12 state basis in the variational approach. 

From the previous example one can see that to get a correct spin-orbit contribution to the energy of a state, it is not enough to include just a few terms in the sum in Eq. \eqref{eq:non degenerate perturbation}. Instead we employ the operators $H^{op}$ given in Eqs. (\ref{eq:operators op},\ref{eq:operators2}). To get a contribution for a 
 particular state, say $|i\rangle$, we apply the 
 L\"owdin perturbation theory.\cite{lowdin1951:JCP} 
For this one has to identify states $|j\rangle$ which are degenerate with $|i\rangle$ with 
respect to the perturbation $H_{SO}$ and which have to be treated exactly.
The rest of the states can be treated perturbatively. The condition for a degeneracy of two states can be taken as $|E^{(0)}_i-E^{(0)}_j| \lesssim \alpha_{SO},\alpha_\mathrm{lin}$, when one considers linear and cubic terms respectively. The finite set of the
 degenerate states will be denoted by $\mathcal{N}$. The effective Hamiltonian
$H^\mathrm{eff}$ acting in $\mathcal{N}$ is
\begin{eqnarray} \label{eq:Lowdin perturbation}
&&H^\mathrm{eff}_{ij}=(H_0+H_{SO})_{ij}+\nonumber\\
&&+\frac{1}{2}\sum_{k\notin\mathcal{N}}
[\frac{(H_{SO})_{ik}(H_{SO})_{kj}}{E^{(0)}_i-E_k^{(0)}}+\frac{(H_{SO})_{ik}(H_{SO})_{kj}}{E_j^{(0)}-E_k^{(0)}}].\qquad
 \end{eqnarray}
For the example of the linear Dresselhaus term, we can now use Eq. \eqref{eq:sum rule} and \eqref{eq:commutators with op} to obtain
\begin{equation} \label{eq:effective hamiltonian}
H^\mathrm{eff}_{ij}=(H_0+H_D)_{ij} - \frac{1}{2}\alpha_D^2 E_0 \left( 1 - \sigma_z l_z \right )_{ij} + R_{ij}, 
\end{equation}
where 
\begin{equation}
R_{ij}=\frac{1}{2}\langle i|H_D P_\mathcal{N} H^{op}_D-H^{op}_D P_\mathcal{N} H_D|j\rangle.
\end{equation} 

First we note that existence of the operator $H_D^{op}$ means that the coupling through $H_D$ between any two states is always much smaller then the difference of the unperturbed energies of these two states, since $(H_D)_{ij}=(E^{(0)}_i-E^{(0)}_j)(H_D^{op})_{ij}\sim (E^{(0)}_i-E^{(0)}_j)\alpha_D$. Then one can partially diagonalize the effective Hamiltonian to eliminate the off-diagonal $H_D$ terms. It turns out, that this leads to a cancellation of the terms $H_D$ and $R$. The effective Hamiltonian is then
\begin{equation} \label{eq:effective hamiltonian2}
H^\mathrm{eff}_{ij}=(H_0)_{ij} - \frac{1}{2}\alpha_D^2 E_0 \left( 1 - \sigma_z l_z \right )_{ij}. 
\end{equation}
This completes the way to get Eq. \eqref{eq:effective} using L\"owdin perturbation theory. There are no linear effects on the double dot energy spectrum from linear spin-orbit terms, which explains the absence of spin hot spots even though symmetry allows that.

The spin-orbit interaction can influence the energy only through the operator $l_z$, which is of the representation $\Gamma_3$, from where we get selection rule--the allowed coupling is between functions of representations $\Gamma_1$ -- $\Gamma_3$ and $\Gamma_2$--$\Gamma_4$. Looking at Fig. \ref{fig:spectrum}, accidental degeneracies of states with such representations are not present in the lower part of the spectrum. The crossing of $\Gamma_1^{21}$ with $\Gamma_4^{11}$ considered in the discussion to Fig. \ref{fig:spectrum} also does not follow the selection rule, hence why the anti-crossing is of the third order. From the selection rules one can immediately see that also the expectation value of $l_z$ is zero in any state.
This result is more general and holds also if the symmetry of the potential is lower (or none), since it follows from the fact that the Hamiltonian $H_0$ is real, so one can choose eigenfunctions to be real. Then the expectation value of any imaginary operator, such as $l_z$, must vanish. We conclude, that apart from degeneracies following from the single dot [that is limits $d\to0(\infty)$] and possible accidental degeneracies respecting the selection rule, double dot states are described by an effective Hamiltonian
\begin{equation} \label{eq:one dim heff}
H^\mathrm{eff}_{ii}=E^{(0)}_i-\frac{1}{2}E_0\alpha_D^2.
\end{equation}
Particularly, the energies of the two lowest states are given by this equation, with an exception for the state $\Gamma_A$ in the region of small $d$ where it is coupled to $\Gamma_4^{11}$ through $l_z$ and we have to describe it here by a $2\times2$ effective Hamiltonian. 

\begin{figure}
\centerline{\psfig{file=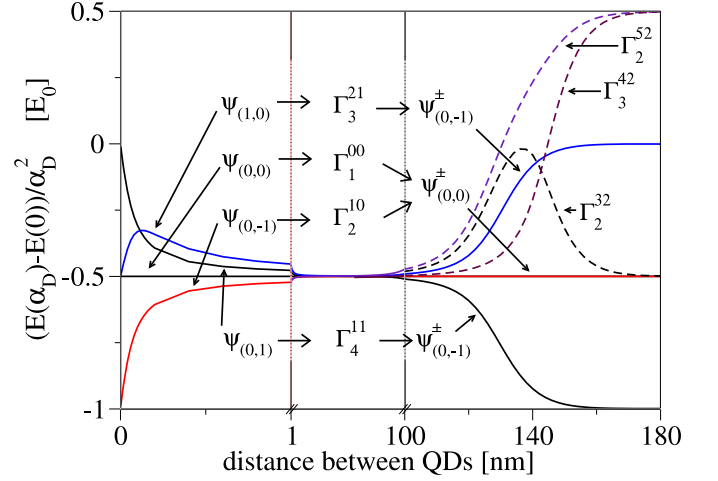,width=0.85\linewidth,angle=270}}
\caption{Calculated corrections to selected lowest energy levels due to $H_D$. All states
have spin $\sigma=+1$. The graph demonstrates a transition between
the symmetry group of the double dot $H_0$ (states $\Gamma$) and that of
single dots (states $\Psi$). 
The transition is induced by $l_z$ which by symmetry couples states $\Gamma_1\leftrightarrow\Gamma_3$ and $\Gamma_2\leftrightarrow\Gamma_4$.
Thus the anti-crossing mechanism will induce transition
$\Gamma_{1(3)}\leftrightarrow \Gamma_1\pm\Gamma_3$ and $\Gamma_{2(4)}\leftrightarrow \Gamma_2\pm\Gamma_4$. These linear combinations are equal to a single dot solution $\Psi_{n,l,\sigma}$ in the case $d\to 0$ and a combination $\Psi_{n,l,\sigma}^\pm\equiv\Psi_{n,l,\sigma}^d\pm\Psi_{n,l,\sigma}^{-d}$ of functions with the same orbital momenta in the case
$d\to\infty$.}
\label{fig:basis transition}
\end{figure}

An illustration of the $l_z$ influence on the spectrum is in Fig. \ref{fig:basis transition}, where the linear Dresselhaus spin-orbit contribution to the energy for several states as a function of the interdot distance is shown.
One can see at what interdot distances the $l_z$ operator causes the qualitative change between the double dot case (where the functions are characterized by a definite representation $\Gamma_i$ and the energy contribution from the spin-orbit is a uniform shift) and the single dot case (where the functions are numbered according to the orbital momentum and the  spin-orbit contribution to the energy depends on $\sigma_z l_z$).  This happens when $E_0\alpha_D^2$ is comparable to the energy difference of the nearly degenerate states. If the criterion for the coupling between the dots is
the constant contribution, $-\alpha_D^2 E_0/2$, to the energy, then 
the double dot region, as far as the spin-orbit coupling is concerned, is between 1 to 100 nm, that is up to
5 times of the confinement length of 20 nm. As an example, for the function $\Gamma_4^{11}$ the coupling in the effective Hamiltonian through $l_z$ to $\Gamma_2^{31}$ is equal to the unperturbed energy difference if $\alpha_D^2\sim d^3e^{-d^2}$, giving $d\approx 3$, corresponding to the interdot distance of $6l_0$. Due to the exponential, this result is insensitive to $\alpha_D$.

The Bychkov-Rashba term can be treated analogously. The effective Hamiltonian is $H^\mathrm{eff}_{ij}=[H_0 - (\alpha_{BR}^2/2) E_0
(1+ \sigma_z l_z)]_{ij}$. The absence of a linear influence on the energy was based on the existence of $H_D^{op}$. Since we found a case where $H_{D3}$ causes linear anti-crossing (see discussion to Fig. \ref{fig:spectrum}), it follows that $H_{D3}^{op}$ can not exist for our double dot potential. However, if one approximates $E_i-E_k\approx E_j-E_k$ in \eqref{eq:Lowdin perturbation}, one can use $H_D^{op}$ to simplify the mixed $D$-$D3$ 
correction. This, according to Fig. \ref{fig:tunneling corrections}, 
is the dominant spin-orbit correction for
the tunneling energy $\delta E_t$. One gets an analogous expression as Eq. \eqref{eq:effective hamiltonian}, where the needed commutator is stated in Eq. \eqref{eq:dd3 commutator}. 
Concluding, if we neglect the mixed $D3$-$D3$ term, we can write the spin-orbit contribution to the energy for the
lowest orbital states to be ($i=S,A$)
\begin{equation} \label{eq:full SO correction}
\delta E_i^{SO}=-E_0(\alpha_D^2+\alpha_{BR}^2)/2+\alpha_D \alpha_{D3} E_0 l_0^2 
(\textbf{k}^2)_{ii}.
\end{equation}

One note to the eigenfunctions: The matrix elements of the
effective Hamiltonian are computed using the eigenfunctions of $H_0$. But the
functions that correspond to the solutions are transformed by
the same unitary transformation that leads from $H_0$ to $H^\mathrm{eff}$. The sum rule
can be used also here to express the influence of $H_{\mathrm{lin}}$ on the
eigenfunctions of $H_0$. If $H_0\Gamma_i=E_i \Gamma_i$, the eigenfunctions corresponding to the effective Hamiltonian, Eq. \eqref{eq:effective hamiltonian}, are
\begin{equation}\label{eq:eigenfunction}
\delta \Gamma_i=\sum_{j\notin \mathcal{N}}\frac{(H_{\mathrm{lin}})_{ji}}{E_i-E_j}\Gamma_j=
-(\mathbb{I}-P_\mathcal{N})H_{\mathrm{lin}}^{op}\Gamma_i.
\end{equation}
Partial diagonalization of the effective Hamiltonian, to go from Eq. \eqref{eq:effective hamiltonian} to Eq. \eqref{eq:effective hamiltonian2}, means we finish the unitary transformation completely and get
$\tilde{\Gamma}_i^\sigma=\Gamma_i^\sigma-H_{\mathrm{lin}}^{op}\Gamma_i^\sigma$ for the eigenfunctions corresponding to the effective Hamiltonian, Eq. \eqref{eq:effective hamiltonian2}.

\subsection{Finite magnetic field.}

The presence of a magnetic field lowers the symmetry of the Hamiltonian without spin-orbit terms. 
The only nontrivial symmetry operator is the inversion $I$ (see Tab. \ref{tab:representations}). 
As a consequence the double dot states fall
into two groups (representations of $C_2$): $\Gamma_1$ and $\Gamma_3$ become $\Gamma_S$ (symmetric under $I$) and 
$\Gamma_2$ and $\Gamma_4$ become $\Gamma_A$ (antisymmetric under $I$). 
Symmetrized functions $g_i^{n,l,\sigma}$ now are
\begin{eqnarray}
g_i^{n,l,\sigma}=\Psi_{n,l,\sigma}^{-d}+D_i\Psi^{d}_{n,l,\sigma}, 
\end{eqnarray}
where the irreducible states $i=S$ and $A$, while the permutation coefficients
$D_S=-D_A=1$. The shifted single-dot wave functions acquire a phase:
\begin{eqnarray}
 \Psi_{n,l,\sigma}^{\pm d}(x,y)=\Psi_{n,l,\sigma}(x\pm l_0 d,y)e^{\pm i d l_0 \theta y/l_B^2},
\end{eqnarray}
depending on which dot they are located.  

\begin{figure}
\centerline{\psfig{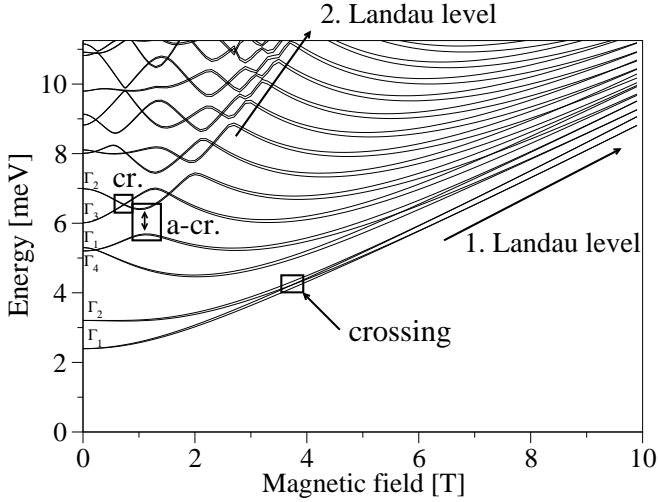}}
\caption{Computed energy spectrum of the double dot Hamiltonian without $H_{SO}$, 
as a function of magnetic field. The quantum dot separation is 50 nm (single dot confinement
length is 20 nm). The energy levels are labeled according to the symmetry of the states
at $B=0$. Two crossings (one between $\Gamma_2$ and $\Gamma_3$, the other between 
$\Gamma_1$ and $\Gamma_2$) and one anti-crossing (between $\Gamma_2$ and $\Gamma_4$) 
are indicated. In the limit $B\to\infty$ the states merge to Landau levels.}
\label{fig:FD for DD}\end{figure}

The double dot energy spectrum of $H_0$ as a function of magnetic field is shown 
in Fig. \ref{fig:FD for DD} for the interdot distance of 50 nm. 
Indicated are two crossings and one anti-crossing induced by magnetic field. 
The first crossing is between $\Gamma_1$ and $\Gamma_2$ (notation from the $B=0$ case).
These two states have opposite spins so they are not coupled and there is no level
repulsion here. (We will see in the next section that spin-orbit coupling will induce
anti-crossing in this case.) The second crossing is between $\Gamma_2$ and $\Gamma_3$,
which behave differently under $I$ and so they are not coupled by magnetic field. 
The actual anti-crossing is between $\Gamma_2$ and $\Gamma_4$, which are both antisymmetric
under $I$. This is an example of anti-crossing induced by magnetic field.

In analogy with Eq. \eqref{eq:full expressions no B} we derive analytical expressions for the energies of the lowest symmetric and antisymmetric states
in the presence of magnetic field using the LCSDO approximation:
\begin{eqnarray}
E_S^{(0)}&\!\!=\!\!&\frac{2E_0}{\eta^2}\Big(\frac{1+[1-d\eta(1-\theta^2)/\sqrt{\pi}]e^{
-(d\eta)^2(1+\theta^2)}}{1+e^{-(d\eta)^2(1+\theta^2)}}-\nonumber\\
&&\quad-\frac{d\eta(1-\theta^2)[e^{-(d\eta)^2}/\sqrt{\pi}-d\eta\,
\mathrm{Erfc}(d\eta)]}{1+e^{-(d\eta)^2(1+\theta^2)}}\Big),\nonumber\\
E_A^{(0)}&\!\!=\!\!&\frac{2E_0}{\eta^2}\Big(\frac{1-[1-d\eta(1-\theta^2)/\sqrt{\pi}]e^{-(d\eta)^2(1+\theta^2)}}
{1-e^{-(d\eta)^2(1+\theta^2)}}-\nonumber\\
&&\quad-\frac{d\eta(1-\theta^2)[e^{-(d\eta)^2}/\sqrt{\pi}-d\eta\,\mathrm{Erfc}(d\eta)]} {1-e^{-(d\eta)^2(1+\theta^2)}}\Big)\!.
\label{eq:full expressions}
\end{eqnarray} 
Here $\eta=l_0/l_B=(1-\theta^2)^{1/4}$. In the limit $d\to\infty$, we can then deduce the tunneling energy in the leading order to be
\begin{eqnarray}
\delta E_t^{(0)} = E_0\frac{4}{\sqrt{\pi}} (1-\theta^2)^{5/4}
 e^{-d^2(1+\theta^2)/\sqrt{1-\theta^2}}.
\end{eqnarray}
If $\theta = 0$, the above expressions reduce to Eq. (\ref{eq:full expressions no B},\ref{eq:dEt}). On the other hand, if $B\to\infty$, then $\delta E_t^{(0)} \sim B^{-5/2}
 e^{-B/B_0}$. 
Magnetic field suppresses $\delta E_t^{(0)}$ by suppressing overlap integrals 
$\langle \Psi^{-d} |H|\Psi^d\rangle$. There are three different effects that magnetic
field introduces. First, the wave functions are squeezed by the confinement 
provided by the vector potential. The natural confinement length is 
$l_B=l_0(1-\theta^2)^{1/4}$, present in the exponential decay factors.
Second, the gauge phase produces factors $(1+\theta^2)$  in 
the exponents of the scalar products $\langle\Psi^{-d}|\Psi^d\rangle$.
Third, as $B$ increases, the confinement potential $V_C$ becomes less
important compared to the confinement of the magnetic vector potential. This
gives rise to the factor $(1-\theta^2)$. Note that in the limit $B\to\infty$,
$\Psi^d_{n,l,\sigma}$ is an eigenstate of $H_0$ for any $d$.
For reasons explained in last paragraph in the Sec. \ref{uncorresponding LCSDO}, Eqs. \eqref{eq:full expressions}
are correct in the limit $d\to \infty$. 
As $d\to 0$, only $E_S^{(0)}$ is correct. The limit value of $E_A^{(0)}(d=0)/E_0$ is $4/(1+\theta^2)$, instead of the exact value of $2(2-\theta)$.

At a finite magnetic field we have also a new term in the Hamiltonian, the Zeeman term.
Since it commutes with $H_0$ the only consequence of this term is a shift of the energy of the states by a value $\sigma\mu_B B$ according to their spin $\sigma$.
Therefore it introduces new crossings of the states with opposite spin. 
An example of this can be seen in Fig. \ref{fig:four lowest no SO}, 
where we plot energies of the four lowest states in the region where the Zeeman shift is comparable to the energy differences of the considered states.  

\subsection{Effective spin-orbit Hamiltonian}

We now study the influence of spin-orbit coupling on the spectrum of double dots in a finite magnetic field. We will
see that spin-orbit terms lead to new spin hot spots even at magnetic fields of the order
of 1 T, and that linear spin-orbit terms will influence tunneling in the second order.
  
Although the presence of the Zeeman term complicates the analysis of the perturbation
theory using operators $H^{op}$, one can still apply the previously developed formalism
if the Zeeman term is treated as a part of perturbation. 
(For a harmonic potential describing
single dots, operators $H_{\mathrm{lin}}^{op}$ have been derived\cite{rodriguez2004b:PRB} for the case of finite magnetic field, so that
the Zeeman term can be included into $H_0$). 
Up to the second order in the perturbation couplings (being now $\alpha_{SO}$ and $\alpha_Z$), 
there is no coupled Zeeman-spin-orbit term. 
This means that in the effective Hamiltonians $H^\mathrm{eff}$ that we already derived for the case of zero magnetic field, 
the Zeeman term appears as a shift of the energies on the diagonal without bringing any new couplings (non-diagonal terms). 
But an important consequence is that the shift can change the number of states we have to include into the basis where the effective Hamiltonian acts, 
because their energy difference to the considered state is comparable to the spin-orbit coupling.

First, in analogy with Eq. \eqref{eq:one dim heff}, if the energy of a state is far enough from others, 
we can consider the basis to consist of one term only and the spin-orbit correction to the energy of state $|i\rangle$ is 
\begin{eqnarray}
\delta E_i^{SO}&=&-E_0\frac{\alpha_D^2}{2}(1-\sigma\overline{L_z})-E_0
\frac{\alpha_{BR}^2}{2}(1+\sigma\overline{L_z})\nonumber\\
&&\qquad+\overline{[H_D^{op},H_{D3}]},
\label{eq:full SO correction with B}
\end{eqnarray}
where the averaging means the expectation value in the state $|i\rangle$ and $\sigma$ is the spin of the state.
Since the presence of magnetic field lowers the symmetry, the last commutator, [Eq. \eqref{eq:dd3 commutator}], can not be simplified according to the
symmetry as was the case before in Eq. \eqref{eq:full SO correction}, and, more important, we no longer have  $\overline{L_z}=0$. 
As a result, there are now corrections to
the tunneling that are of the second order in the linear spin-orbit couplings. 
These corrections depend on $\alpha_-^{(2)}\equiv \alpha_D^2-\alpha_{BR}^2$.

\begin{figure}
\centerline{\psfig{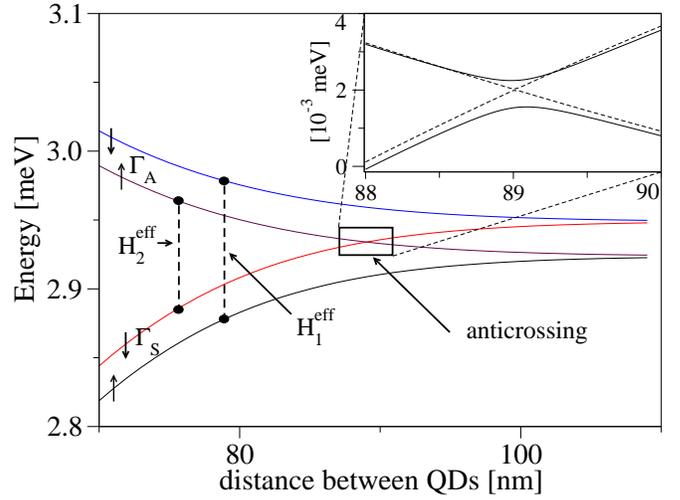}}
\caption{Calculated energies of the four lowest states of Hamiltonian
$H_0+H_Z$ at $B=1$ T. The vertical dashed lines indicated doublets in the effective Hamiltonian. 
The inset displays the anti-crossing at 89 nm due to $H_D$. 
Dashed lines are energies of $H_0+H_Z$, solid lines of $H_0+H_Z+H_D$.}
\label{fig:four lowest no SO}\end{figure}

Second, we look how the energies of the four lowest states are changed, using again the example of the linear Dresselhaus term. 
They are plotted in Fig. \ref{fig:four lowest no SO}. 
Here in the main figure one can see the shift caused by the Zeeman term and the anti-crossing
 induced by the spin-orbit coupling is magnified in the inset. 
 The anti-crossing states are $\Gamma_S^\downarrow$ and $\Gamma_A^\uparrow$. 
 In the case of zero magnetic field we described each of the four basis states by Eq. \eqref{eq:one dim heff}. Now, in principle, we have to describe them by a $4\times 4$ effective Hamiltonian Eq. \eqref{eq:effective hamiltonian}. Due to symmetry we can simplify this Hamiltonian into two $2\times 2$ Hamiltonians, $H_1^\mathrm{eff}, H_2^\mathrm{eff}$, acting in the bases $\Gamma_S^\uparrow,\Gamma_A^\downarrow$ and $\Gamma_S^\downarrow,\Gamma_A^\uparrow$ respectively. The four components of the effective Hamiltonian matrix are 
\begin{eqnarray}
(H^\mathrm{eff})_{11}&=&E_S^{(0)}-E_0\frac{\alpha_D^2}{2}(1-\sigma( L_z)_{11})-\sigma\mu_B B-R_{11}\nonumber\\
(H^\mathrm{eff})_{22}&=&E_A^{(0)}-E_0\frac{\alpha_D^2}{2}(1+\sigma (L_z)_{22})+\sigma\mu_B B+R_{11}\nonumber\\
(H^\mathrm{eff})_{12}&=&(H^\mathrm{eff})_{21}^\dagger=(H_D)_{12}
\end{eqnarray}
where $\sigma=1$ for $H^\mathrm{eff}_1$ and $\sigma=-1$ for $H^\mathrm{eff}_2$, while indices 1,2 denote the first and the second term in the corresponding basis. 
Comparing to the case of zero magnetic field the Zeeman term increases the difference of 
the diagonal elements in $H^\mathrm{eff}_1$ and decreases them in $H^\mathrm{eff}_2$. 
The ground and the fourth excited states which are described by $H^\mathrm{eff}_1$ stay 
isolated, and we can do the perturbative diagonalization to get rid of the off-diagonals. The energy of the two states is then up to the second order in the spin-orbit coupling accurately described 
by Eq. \eqref{eq:full SO correction with B}. 
Concerning the states $\Gamma_S^\downarrow$ and $\Gamma_A^\uparrow$, there is a region in the 
interdot distance of a few nanometers, where the two states must be described by the two 
dimensional $H_2^\mathrm{eff}$ to account for the anti-crossing, 
which is caused by the the matrix element 
$\langle \Gamma_S^\downarrow| H_D |\Gamma_A^\uparrow\rangle$. 
LCSDO gives for this element a result correct only in the order of magnitude. 
This is because even in the limit $d\to\infty$ this matrix element is of the same order as the 
neglected coefficients $c(n,l)$ in the LCSDO approximation, Eq. \eqref{eq:LCSDO}.

\begin{figure}
\centerline{\psfig{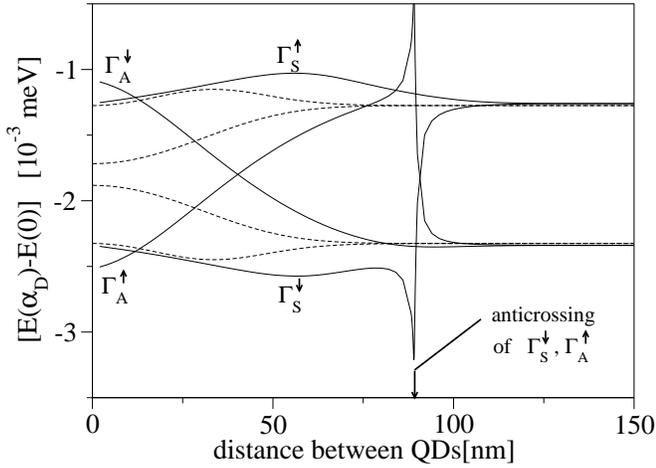}}
\caption{Calculated corrections to the energies of the four lowest states due to the
linear Dresselhaus term $H_D$, at $B=1$ T. Solid lines are numerical data, dashed
lines are analytical expressions computed by Eq. \eqref{eq:full SO correction with B} using 
the LCSDO approximation for the states.}
\label{fig:SO corrections}\end{figure}

The spin-orbit corrections to the energies from $H_D$ for
the four lowest states as functions of the interdot distance are in
Fig. \ref{fig:SO corrections}. Also shown are analytical values computed by Eq. 
\eqref{eq:full SO correction with B}, that is, ignoring anti-crossing. The scale implies that the
corrections are of the second order in $\alpha_D$ and for the states $\Gamma_S^\downarrow$ and $\Gamma_A^\uparrow$ are enhanced in the anti-crossing
region. 
The incorrect values at the limit $d\to0$ for functions $\Gamma_A^\sigma$ are because of 
reasons explained in the last paragraphs in Sec. \ref{uncorresponding LCSDO}.

\subsection{Spin-orbit corrections to the effective $g$-factor and tunneling frequency}

We next analyze spin-orbit corrections to the $g$-factor, 
$\delta g\equiv[\delta E(\Gamma_S^\downarrow)-\delta E(\Gamma_S^\uparrow)]/\mu_B B$,
that characterizes the energy cost of a spin flip in the ground state, 
or the frequency of a spin precession.
Another kind of oscillation is electron tunneling,
when electron oscillates between the left and the right dot without changing its spin. 
The frequency of this oscillation, $\delta E_t/2\hbar$, is given by the energy difference $\delta E_t= E(\Gamma_A^\uparrow)-E(\Gamma_S^\uparrow)$. 
Corrections to this energy difference induced by the spin-orbit interaction are denoted in this paper as $\delta E_t^{SO}$. 

\begin{figure}
\centerline{\psfig{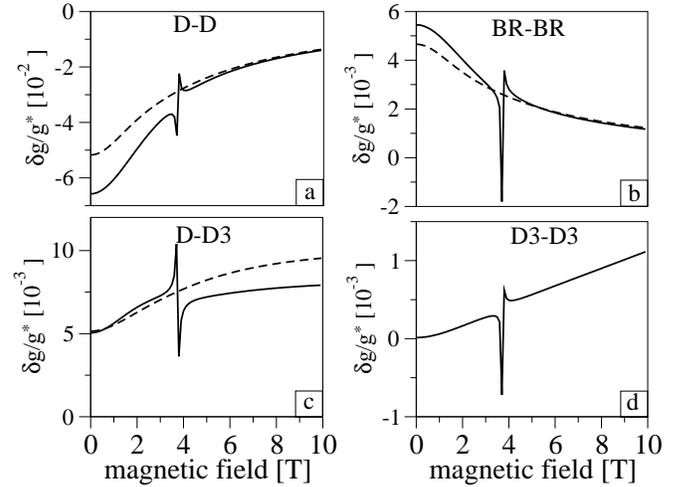}}
\caption{Calculated spin-orbit corrections (relative to the conduction band value $g^*$) to the effective $g$-factor, $\delta g=[\delta E(\Gamma_S^\downarrow) - \delta E (\Gamma_S^\uparrow)]/\mu_B B$,
as functions of magnetic field. The distance between the dots is $50$ nm. 
Solid lines are numerical data, dashed lines are analytical values computed using Eq. \eqref{eq:full SO correction with B}. 
Contributions come from linear spin-orbit 
terms (a, b), the mixed Dresselhaus correction
from  $H_D$ and $H_{D3}$ (c) and the cubic Dresselhaus $H_{D3}$ correction (d).}
\label{fig:cors to deg}\end{figure}

First, we take a look at the spin-orbit corrections to the $g$-factor. Contributions in the second order of the spin-orbit couplings are shown in Fig. \ref{fig:cors to deg},
 as functions of magnetic field at a constant interdot distance. The spin-orbit 
contribution to the $g$-factor in the double dot case has the same qualitative dependence on 
the magnetic field as in the single dot case (see discussion to Fig. \ref{fig:g factor corrections}). 
However, at finite interdot distances, there is an enhancing effect on the 
spin-orbit contributions. This can be seen in Fig. \ref{fig:SO corrections}, where at a certain 
magnetic field, the spin-orbit contribution to the $g$-factor is enhanced for a finite $d$ 
compared to the case of $d=0(\infty)$. We found numerically, that the enhancement can be up to 
50\% of the value of the correction in $d=0$ at magnetic fields of the order of 1 T. 

At the anti-crossing the spin-orbit contributions show cusps. At magnetic fields bellow the anti-crossing, the dominant spin-orbit contribution is $D$-$D$ which reduces the conduction band $g$-factor by several percent. Contributions $D$-$D3$ and $BR$-$BR$ are 
one order of magnitude smaller. Using Eq. \eqref{eq:full SO correction with B}, that is 
ignoring the anti-crossing, we get for the contribution from the linear spin-orbit terms 
\begin{equation}
\delta g(\mathrm{lin}-\mathrm{lin})=-\frac{E_0}{\mu_B B}\alpha_-^{(2)}\overline{L_z},
\end{equation}
where, in the limit $d\to\infty$,
\begin{eqnarray}
\overline{L_z}\approx\theta[1+(d/\eta)^2e^{-(d\eta)^2(1+\theta^2)}].
\end{eqnarray}
From Fig. \ref{fig:cors to deg} one can see that the analytical result agrees with numerics. 

\begin{figure}
\centerline{\psfig{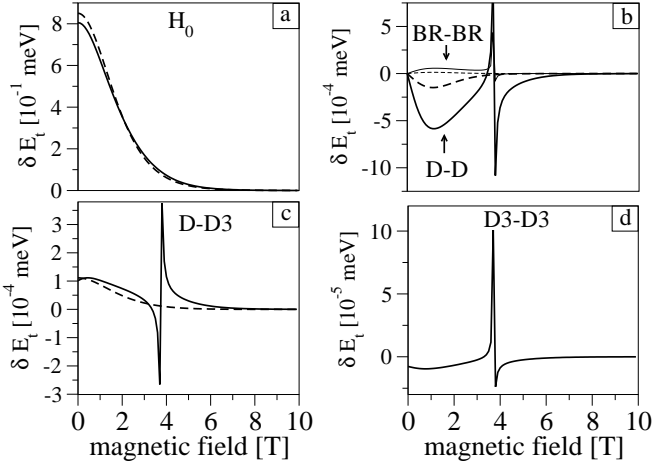}}
\caption{Calculated spin-orbit corrections to the tunneling energy $\delta E_t$ as a function
of magnetic field. The interdot distance is 50 nm. Solid lines are numerical data, 
dashed lines are analytical expressions computed by Eq. \eqref{eq:full SO correction with B}.}
\label{fig:cors to tun} \end{figure}

Finally, we look at the tunneling energy in the presence of both magnetic field and spin-orbit
couplings. The spin-orbit corrections, as a function of magnetic field, are shown in 
Fig. \ref{fig:cors to tun}. At zero magnetic field there is no contribution from the linear 
terms (result of the section \ref{sec:DDSOC1}) and the dominant contribution is $D$-$D3$. 
Similarly to $\delta E_t^{(0)}$, the corrections decay exponentially
with increasing magnetic field. Anti-crossing strongly influences the tunneling energy. Using
LCSDO for $d\to\infty$ we obtain in the leading order
\begin{equation}
\label{eq:tunneling correction}
\delta E_t^\mathrm{lin} = -E_0\alpha^{(2)} \theta (d/\eta)^2e^{-(d\eta)^2(1+\theta^2)}.
\end{equation}
This analytical formula underestimates the corrections from the linear spin-orbit terms by a 
factor of $\sim 3$. Nevertheless the analytical expression for 
$D$--$D3$ is reasonably good. In the magnetic field below anti-crossing, the relative change of the tunneling energy stemming from the spin-orbit terms is of 
order $10^{-3}$.

\subsection{Tunneling Hamiltonian}

We now use our results to describe the influence of the spin-orbit interaction on the lower part of the spectrum. We restrict our Hilbert space on the four lowest states $\Gamma_{S(A)}^\sigma$, the eigenstates of the total double dot Hamiltonian. 
Because of the transformation Eq. \eqref{eq:eigenfunction} these four states have neither definite 
symmetry with respect to inversion $I$, nor a definite spin in z-direction.
In this section we will denote them as spin `up' and spin `down' states.
For description of a transport through the double dot it is useful to define the following 
left and right localized functions 
\begin{equation}
L^\sigma(R^\sigma)=\frac{1}{\sqrt{2}}(\Gamma_S^\sigma \pm \Gamma_A^\sigma),
\end{equation}
where plus and minus holds for $L$ and $R$, respectively. In the limit $d\to\infty$ these 
functions converge to single dot solutions localized in the left and 
right dot. 

The effective Hamiltonian of our system in the second quantization formalism is 
\begin{equation}
\label{eq:second quantization}
H=\sum_{\sigma=\uparrow,\downarrow} E^\sigma(n_{L\sigma}+n_{R\sigma})-
(\delta E_t^\sigma/2)(a^\dagger_{L\sigma}a_{R\sigma}+a^\dagger_{R\sigma}a_{L\sigma}),
\end{equation}
where $E^\sigma=(E_S^\sigma+E_A^\sigma)/2$, $\delta E_t^\sigma=E_A^\sigma-E_S^\sigma$, while 
$a^\dagger$, and $a$ are creation and annihilation operators, and $n=a^\dagger a$. We can get both localized and 
spin-pure states if we diagonalize $\sigma_z$ in a chosen basis. 
For example, taking $L^\uparrow$ and $L^\downarrow$,  
we get  $L^{\text{pure}\uparrow}\sim (L^\uparrow+oL^\downarrow)$ and 
$L^{\text{pure}\downarrow}\sim (L^\downarrow-o^+L^\uparrow)$, up to normalization 
$(1-|o|^2/2)$. That is, the left pure spin state is a linear superposition of both left states 
with spin `up' and `down'. The coefficient $o$ is proportional to $\alpha_{SO}$. 

In the following we are interested in the time evolution of localized states given by 
Hamiltonian Eq. \eqref{eq:second quantization}. First we note, that due to the non-diagonal 
terms, the electron which is in a localized state will tunnel into the other localized state 
after the tunneling time $t_\mathrm{tun}^\sigma=h/\delta E_t^\sigma$, resulting in coherent 
oscillations. For our parameters $t_\mathrm{tun}\approx 1$ ps. In the Hamiltonian there is no mixing between spin `up' and `down' states. 
However, there will be mixing (or spin-flip) if we work with localized pure spin states. 
Electron being originally in $L^{\text{pure}\uparrow}$ will, after the tunneling time $t_\mathrm{tun}^\uparrow$, 
contain $R^{\text{pure}\downarrow}$ with the probability amplitude 
\begin{equation} \label{eq:spin flip}
c=io\pi(\delta E_t^\downarrow-\delta E_t^\uparrow)/2\delta E_t^\uparrow,
\end{equation}
assuming that the difference in $\delta E_t$ for different spins is much smaller that $\delta E_t$ itself.

In the case of zero magnetic field, because of Kramer's degeneracy, the tunneling frequencies 
are the same for both spin orientations. Then whatever is the initial combination of spin 
`up' and `down' (let it be, for example, $L^{\text{pure}\uparrow}$), during the time 
evolution (oscillations) there will be no relative change in the coefficients in this linear 
combination. Therefore spin-orbit coupling does not lead to spin-flipping and $c=0$ in Eq. 
\eqref{eq:spin flip}.

\textit{In a finite magnetic field, the tunneling frequency for spin `up' and `down' are different.} 
The difference is caused only by spin-orbit terms, and is of order $\alpha_-^{(2)}=\alpha_D^2-\alpha_{BR}^2$. Equation \eqref{eq:full SO correction with B}
shows, that the contribution to $\delta E_t^\uparrow$ from the linear spin-orbit terms is opposite 
that of $\delta E_t^\downarrow$ and therefore their difference is twice the expression in Eq. 
\eqref{eq:tunneling correction}. We conclude, that spin-flip during tunneling induced by 
spin-orbit coupling is proportional to the third power in spin-orbit couplings and depends 
linearly on the magnetic field if the magnetic field is small ($c\sim \alpha_\mathrm{lin}^3\theta$).

The different tunneling frequency can be exploited for separation of different spin states in a homogeneous magnetic field. 
Starting with some combination of `up' and `down' states localized in one dot, after time 
$t_\mathrm{sep}=h/(\delta E_t^\uparrow-\delta E_t^\downarrow)=t_\mathrm{tun}\delta E_t^\uparrow/(\delta E_t^\uparrow-\delta E_t^\downarrow)$ 
the part with spin `up' will be localized in the left, and the spin `down' will be in the 
right dot. From Fig. \ref{fig:cors to tun} one can see that one needs about $10^3$ 
coherent oscillations to get the spatial spin separation. Therefore the decoherence time must 
be longer in order to observe this effect experimentally. We note that the separated states 
will not be pure spin states, but will contain a small (linearly proportional to 
$\alpha_{SO}$) admixture of opposite pure spin states. 

\section{Conclusions}

We have performed numerically exact calculation of the spectrum of a single electron localized by a confining potential 
in single and double GaAs quantum dots. We have studied the influence of the spin-orbit terms,
namely the Bychkov-Rashba and the linear and cubic Dresselhaus terms, on the energy spectrum. 
In the single dot case we have elaborated on previous results and shown that the spin-orbit interaction has
three principal effects on the spectrum: first, the interaction shifts the 
energy by a value proportional to the second order in the spin-orbit couplings, second, it 
lifts the degeneracy at zero magnetic field, and, third, the Bychkov-Rashba term  gives rise
to spin hot spots at finite magnetic fields. 

In the double dot case we have addressed the symmetries of the Hamiltonian. For zero magnetic 
field without spin-orbit terms we have constructed the correlation diagram, between  single
and double dot states, of the spectrum. We have used properly symmetrized linear combination 
of shifted single dot solutions as an approximation for a double dot solution and found  
that for the four lowest states it gives a good approximation for the energy. 
As for the contributions to the energy from the linear spin-orbit terms, we have found that in zero 
magnetic field a typical feature of a double dot is a uniform shift of the energy 
proportional to the second order in the coupling strengths without any dependence on the 
interdot distance. This is true also if the potential has lower (or none) symmetry (for example biased dots). 
Therefore, in zero magnetic field, there is no influence on the 
tunneling frequency up to the second order in the linear spin-orbit couplings and the 
dominant contribution comes from the mixed linear and cubic Dresselhaus second order term. We found, that spin hot spots in zero magnetic field exist in the double dot, but are solely due to the cubic Dresselhaus term. This means also, that for our potential, for the cubic Dresselhaus term there can not exist a unitary transformation to eliminate its contribution in the first order.

The effective $g$-factor, on the other hand, is influenced by the second order linear 
spin-orbit couplings even at $B\sim 0$, so the dominant contribution here is the linear 
Dresselhaus term for GaAs. In finite magnetic fields the uniform shift does not hold any more and 
there is a contribution to the tunneling frequency in the second order of the 
linear spin-orbit couplings. We have derived 
an effective Hamiltonian, using L\"owdin's perturbation theory, with which analytical results up 
to the second order in perturbations (Zeeman and spin-orbit terms with the exception of cubic 
Dresselhaus-cubic Dresselhaus contribution) can be obtained provided one has exact solutions of 
the double dot Hamiltonian without Zeeman and spin-orbit terms. From this effective 
Hamiltonian we have derived the uniform shift in zero magnetic field. In a
finite magnetic field we used linear combinations of single dot solutions to obtain analytical 
expressions for the spin-orbit contributions to the energy for the four lowest states. We 
have analyzed them as functions of the interdot distance and magnetic field and compared them with 
exact numerical values. The spin-orbit relative contribution to the $g$-factor and the 
tunneling frequency is of the order of $\sim 10^{-2}$ and $\sim 10^{-3}$, respectively. Due to 
the degeneracy of the energy spectrum at large interdot distance the spin hot spots exist 
also at smaller magnetic fields compared to the single dot case.
 
As an application of our results we have constructed an effective Hamiltonian acting in a 
restricted Hilbert space of four states--electron with spin up and down (these are effective
spins in the presence of spin-orbit coupling) localized on 
either dot. Effectively, there is only spin-conserving tunneling between the localized states, no spin-flip tunneling. 
In zero magnetic field the spin-orbit interaction does not 
significantly influence the tunneling frequency, nor it implies spin-flip tunneling. In finite magnetic fields
the tunneling frequency is spin dependent, the difference being of second order in linear spin-orbit terms. This leads to a spin 
flip amplitude proportional to the third power in spin-orbit couplings (it is linear in magnetic field). 
We propose to use this difference of the tunnelings to spatially separate electron spin 
in homogeneous magnetic field.

\acknowledgments

We thank Ulrich Hohenester for usefull discussions.
This work was supported by the US ONR.

\bibliography{../references/quantum_dot,../references/electron-phonon}

\end{document}